\documentclass[graybox]{svjour3}

\usepackage{mathptmx}       
\usepackage{helvet}         
\usepackage{courier}        
\usepackage{type1cm}        
\usepackage{makeidx}         
\usepackage{graphicx}        
\usepackage{multicol}        
\usepackage[bottom]{footmisc}
\usepackage{amsmath, amsfonts, amssymb}
\usepackage{stmaryrd}
\usepackage{color}
\usepackage{tikz}
\usepackage{dune}
\usepackage{todonotes}
\usepackage{url}
\usepackage{doi}
\usepackage{comment}
\usepackage{hyperref}
\hypersetup{pdftitle={The DUNE-Fem-DG framework},
            pdfborderstyle={/S/U/W 1}}

\usepackage{xspace}
\usepackage{bbm}
\usepackage[titletoc]{appendix}
\usepackage{subfig}

\def\norm#1{\|#1\|}

\newcommand{\phispace}{V_h}

\newcommand{\grid}{{\mathcal{T}_h}}
\newcommand{\entity}{E}
\newcommand{\elem}{\entity}
\newcommand{\neig}{K}
\newcommand{\isec}{e}

\newcommand{\RRR}{{\mathbbm R}}
\newcommand{\NNN}{{\mathbbm N}}

\newcommand{\Fc}{F_c}
\newcommand{\Fv}{F_v}
\newcommand{\Simpl}{S_{i}}
\newcommand{\Sexpl}{S_{e}}

\newcommand{\flux}[1]{\widehat{#1}}

\newcommand{\fluxF}{\flux{F_c}}
\newcommand{\fluxA}{\flux{F_v}}

\newcommand{\vect}[1]{\boldsymbol{#1}}
\newcommand{\vecU}{\vect{U}}
\newcommand{\vecV}{\vect{V}}

\newcommand{\vecv}{\vect{v}}

\newcommand{\basefct}{\vect{\varphi}}

\newcommand{\sol}{\vecU}
\newcommand{\df}{\sol_h}
\newcommand{\uleft}{\sol}
\newcommand{\uright}{\vecV}
\newcommand{\um}{\uleft}
\newcommand{\up}{\uright}
\newcommand{\ubar}{\bar{\sol}}

\newcommand{\neigh}{\oper{N}}

\newcommand{\oper}[1]{\mathcal{#1}}
\newcommand{\spcoper}{\oper{L}_h}
\newcommand{\limiter}{\Pi_h}
\newcommand{\limitedoper}{\widetilde{\oper{L}}_h}

\newcommand{\vjump}[1]{[ \! [ {#1} ] \! ]_{\isec} }

\newcommand{\vaver}[1]{\{ \! \! \{ {#1} \} \! \! \}_{\isec} }

\newcommand{\su}{S(\df)}
\newcommand{\indicator}{J}

\newcommand{\n}{\vect{n}}
\renewcommand{\ne}{\n_{\isec}}
\newcommand{\interset}{\mathcal{I}_{\elem}}
\newcommand{\dfelem}{\sol_{\elem}}
\newcommand{\dfneig}{\sol_{\neig}}
\newcommand{\ubarneig}{\bar{\sol}_{\neig}}

\makeindex             

\newcommand{\suminter}[2]{ \mbox{\scriptsize$
                            \renewcommand{\arraystretch}{0.5}
                            \begin{array}{c}
                            \isec \in #1, \\
                            \vecv \cdot \n_{\isec} #2 0
                        \end{array}$
                        }
                        }

\def\makeheadbox{}                        

\begin{document}

\title{Extendible and Efficient Python Framework for Solving Evolution
Equations with Stabilized Discontinuous Galerkin Methods} 
\titlerunning{Python Framework for Solving Evolution Equations with Stabilized DG}
\author{Andreas Dedner, Robert Kl\"ofkorn$^*$}

\institute{
Andreas Dedner 
\at 
University of Warwick, UK \\
\email{A.S.Dedner@warwick.ac.uk}
\and 
Robert Kl\"ofkorn ($^*$corresponding author)
\at 
Lund University, Box 118, 22100 Lund, Sweden
\\
\email{robertk@math.lu.se} 
}

\date{}

\maketitle

\abstract{
This paper discusses a Python interface for the recently published 
\dunefemdg module which provides highly efficient implementations 
of the Discontinuous Galerkin (DG) method for solving a wide 
range of non linear partial differential equations (PDE). 
Although the C++ interfaces of \dunefemdg are highly flexible and customizable, 
a solid knowledge of C++ is necessary to make use of this powerful tool. 
With this work easier user interfaces based on Python and the Unified Form
Language are provided to open \dunefemdg for a broader audience. 
The Python interfaces are demonstrated for both parabolic and first order hyperbolic PDEs.}
\keywords{\dune, \dunefem, Discontinuous Galerkin, Finite Volume, Python, 
Advection-Diffusion, Euler, Navier-Stokes
\\[5pt]
{\bf MSC }(2010){\bf:} 
65M08, 
65M60, 
35Q31, 
35Q90, 
68N99
}

In this paper we introduce a Python layer for the \dunefemdg\footnote{\url{https://gitlab.dune-project.org/dune-fem/dune-fem-dg.git}}
module \cite{dunefemdg:17} which is available open-source. 
The \dunefemdg module is based on \dune~\cite{dunepaperII:08} and \dunefem~\cite{dune:Fem} in particular 
and makes use of the infrastructure implemented by \dunefem  for seamless integration of
parallel-adaptive Finite Element based discretization methods.

\dunefemdg focuses exclusively on
Discontinuous Galerkin (DG) methods for various types of problems. The discretizations used
in this module are described by two main papers, \cite{limiter:11} where we introduced
a generic stabilization for convection dominated problems that works on generally
unstructured and non-conforming grids and \cite{cdg2:12} where we
introduced a parameter independent DG flux discretization for diffusive operators.
\dunefemdg has been used in several applications (see \cite{dunefemdg:17} for a
detailed list), most notably a comparison with the production code of the German Weather Service COSMO
has been carried out for test cases for atmospheric flow
(\cite{dunecosmo:12,cosmodg:14}).
The focus of the implementation is on Runge-Kutta DG methods using mainly
a matrix-free approach to handle implicit time discretizations which is a
method especially used for convection dominated problems.

Many software packages provide implementations of DG methods, for example, 
deal.II~(\cite{BangerthHartmannKanschat2007}), feel++~(\cite{feelpp}),
Nektar++~(\cite{ka.sh:05}), FLEXI~(\cite{flexi:12}),  Fenics~(\cite{fenics}),
or Firedrake~(\cite{firedrake}). 
However, most of these packages do not combine the complete set of the following features:
\begin{itemize}
\item a wide range of different DG schemes for time dependent diffusion,
      advection-diffusion, and purely hyperbolic problems;
\item a variety of grids ranging from dedicated Cartesian to 
   fully unstructured in $2$, and $3$ space dimensions;
\item dynamic local refinement and coarsening of the grid;
\item parallel computing capabilities with dynamic load balancing;
\item rapid prototyping using a Python front-end;
\item open-source licenses.
\end{itemize}
All of the above feature are available in the framework described here.
It is based on the \emph{Distributed and Unified Numerics Environment} (\dune),
which provides one of the most flexible and comprehensive grid interfaces
available. The interfaces allows to conveniently switch grid implementations
(not just just element types), without the need to re-write application code. 
Various different implementations support all kinds of grid structures from Cartesian grids
to polyhedral grids, all with their own optimized data structure and implementation.
This is a very unique feature of the \dune framework, which is also
available in the \dunefemdg package described in this paper.

So far, a shortcoming of \dunefemdg has been the template heavy and relatively complicated C++ user
interfaces, leading to a steep learning curve for implementing new models and 
applications or coupling of such. To simplify the usage of the software
especially for new users, recent development (e.g. \cite{fvca9:20}) have focused on
adding a Python layer on top of \dunefemdg. Low level Python bindings were introduced for 
the \dune grid interface in \cite{dune-python} and a detailed tutorial
providing high level access to \dunefem is also available \cite{dunefempy}.
The software can now be completely installed from the \emph{Python Package Index}
and used from within Python without requiring the users to change and compile any of
the C++ code themselves. 
The bindings are setup so that the flexibility of the \dune framework is
not compromised, while at the same time providing a high level abstraction
suitable for rapid prototyping of new methods.
In addition a wide range of mathematical
models can be easily investigated with this software which can be described
symbolically using the domain specific Unified Form Language (UFL) \cite{UFL}.

The Fenics project pioneered the Unified From Language for the convenient 
formulation of weak forms of PDEs. Over time other finite element frameworks, for example,
firedrake \cite{firedrake}
or ExaStencils (\cite{exastencils:20}) adopted UFL to transform the weak forms into C or C++ code that assembles 
resulting linear and linearized forms into matrix structures, for example
provided by packages like PETSc~(\cite{petsc-user-ref}). To our knowledge only a few packages
exists based on UFL that consider convection dominated evolution
equations. For example, in \cite{Houston:18} UFL is used to 
describe weak forms for the compressible Euler and Navier-Stokes equations
but only in the stationary setting.
Another example is \cite{wallwork:20} where shallow water applications are considered.
A package that also provides Python bindings (not using UFL) with a strong focus on
hyperbolic problems is \cite{mandli2016clawpack}, where higher order
Finite-Volume schemes are the method of choice.
To our best knowledge this is the first package which combines high level
scripting support with efficient stabilized Discontinuous Galerkin
methods for solving the full range of pure hyperbolic systems, through advection
dominated problems, to diffusion equations.

UFL is flexible enough to describe some DG methods like the interior
penalty method directly. While this is also an option within \dunefemdg,
we focus on a slightly different approach in this paper, where we use
UFL to describe the strong from of the equation, similar to \cite{Houston:18}.
The model description in UFL is then used in combination with 
pre-implemented discretizations of a variety of different DG schemes to
solve the PDE. This approach allows us to also provide DG methods
that can not be easily described within UFL, including 
interesting diffusion discretization like CDG, CDG2, BR2, or even LDG, but also provides
an easier framework to introduce complex numerical fluxes and limiters
for advection dominated problems.
For the time discretization \dunefemdg uses a method of lines approach,
providing the user with a number of Strong Stability Preserving (SSP) implicit, explicit, and IMEX schemes.
Overall the package thus offers not only a strong base for building state of 
the art simulation tools but it also allows for the development of new
methods and comparative studies of different method. In addition, parallelization and
adaptivity, including $hp$ adaptivity, can be used seamlessly with all
available methods.

\subsubsection*{Motivation and aim of this paper}

This paper describes a collaborative effort to establish a test and research
environment for DG methods based on \dune and \dunefem 
for advection-diffusion problems ranging from advection dominated to diffusion only problems. 
The aim here is to provide easy access to a comprehensive 
collection of existing methods and techniques to serve as a starting 
point for new developments and comparisons without having
to reinvent the wheel. This is combined with the easy access to
advanced features like the availability of different grid element types,
not limited to dimensions less than or equal to three,
$hp$ adaptation, and parallel computation with both distributed (e.g. MPI) and 
shared memory (e.g. OpenMP) parallelization, 
with excellent scalability \cite{dgimpl:12} even for large core counts and
very good on-core efficiency in terms of floating point 
operations per second \cite{dunefemdg:17}. 
Python has become a widespread programming language 
in the scientific computing community including the use in industry 
we see great potential in improving the scientific development of DG methods 
by providing access to state of the art research tools which can be used
through the Python scripting language, backed up by an efficient C++
back-end.

The focus of this paper is to describe the different parts of the framework and how they can be
modified by the user in order to tailor the code to their needs and research
interests. In this paper we will mainly look at advection dominated
evolution equations because most diffusion dominated problems can be often
completely described within a variational framework for which the domain
specific language UFL is very well suited and the standard code generation
features available in \dunefem are thus sufficient. For advection dominated
problems additional ingredients are often required, e.g., for
stabilization, that do not fit the variational framework. This aspect of
the algorithm will be a central part of this paper.
The development and improvement of the DG method in general or the high performance computing 
aspect of the underlying C++ framework, which has been investigated 
previously, are outside the scope of this work.

The paper is organized as follows. 
In Section \ref{sec:equations} we briefly recall the main building blocks
of the DG discretization of advection diffusion problems.
In Section \ref{sec:implementation} we introduce the newly developed 
Python based model interface.
In Section \ref{sec:codegen} we investigate the performance impact of using
Python scripting and conclude with discussing the extensibility of the
approach in Section \ref{sec:conclusion}. 
Installation instructions are given in Appendix \ref{sec:installation} and additional code examples are available
in Appendix \ref{sec:extension} and \ref{app:modalind}.


\section{Governing Equations, Discretization, and Stabilization}
\label{sec:equations}
We consider a general class of time dependent nonlinear advection-diffusion-reaction problems
for a vector valued function $\sol\colon(0,T)\times\Omega\to\RRR^r$
with $r\in\mathbb{N}^+$ components of the form
\begin{eqnarray} 
\label{eqn:general}
\label{eqn:ns}
\label{eqn:general_op}
  \partial_t \sol  = \oper{L}(\sol) &:=& - \nabla \cdot\big( \Fc(\sol) -
       \Fv(\sol,\nabla\sol) \big) + \Simpl(\sol) + \Sexpl(\sol) \ \ \mbox{ in } (0,T] \times \Omega
\end{eqnarray}
in $\Omega \subset \RRR^d$, $d=1,2,3$. Suitable initial and boundary conditions have
to be added. $\Fc$ describes the convective flux, $\Fv$ the viscous flux,
$\Simpl$ a stiff source term and $\Sexpl$ a non-stiff source term.
Note that all the coefficients in the partial differential equation are
allowed to depend explicitly on the spatial variable $x$ and on time $t$
but to simplify the presentation we suppress this dependency in our
notation. Also note that any one of these terms is also allowed to be zero.

For the discretization we use a method of lines approach based on first
discretizing the differential operator in space using a DG approximation
and then solving the resulting system of ODEs using a time stepping scheme.


\subsection{Spatial Discretization}
\label{sec:dune}
\label{seq:discretization_spatial}
\newcommand{\dual}[1]{\langle \basefct, #1 \rangle}%
Given a tessellation $\grid$ of the computational domain $\Omega$ with
$\cup_{K \in \grid} K = \Omega$ we introduce a piecewise 
polynomial space 
$    \phispace = \{\vecv\in L^2(\Omega,\RRR^{r}) \; \colon
    \vecv|_{K}\in[\mathcal{P}_k(K)]^{r}, \ K\in\grid\}
      \ \textrm{for some}\;k \in \NNN,$
where $\mathcal{P}_k(K)$ is a space containing all polynomials up to degree $k$.
Furthermore, we
denote with $\Gamma_i$ the set of all intersections between two 
elements of the grid $\grid$ and accordingly with $\Gamma$ the set of all
intersections, also with the boundary of the domain $\Omega$. 

A major advantage of the discontinuous Galerkin methods over other
finite-element methods is that there
is little restriction on the combination of grids and discrete
spaces one can use.
Due to the low regularity requirements of discontinuous Galerkin methods we can use any
form of grid elements, from (axis aligned or general) cubes,
simplices, to general polyhedrons. The best choice will depend on the
application. In addition different local adaptation strategies can
be considered as part of the underlying grid structure, e.g.,
conforming refinement using red-green or bisection strategies, 
or simple nonconforming refinement.

For approximation purposes the discrete spaces are
generally chosen so that they contain the full polynomials of a given
degree on each element. 
But even after having fixed the space,
the actual set of basis functions used to represent the space,
can have a huge impact on the performance of the scheme.
Central here is the structure of the local mass matrix on each element of the
grid. During the time evolution, this has to be inverted in each time step, so efficiency here
can be crucial. So a possible choice is to use a polynomial space
with basis functions orthonormalized over the reference elements in
the grid. If the geometric mapping between physical grid elements and
these reference element is affine then the resulting local mass
matrix is a very simple diagonal matrix. If the mapping is not affine
then the mass matrix depends non trivially on the element under
consideration and will often be dense.
In this case another approach is to use Lagrange type
basis functions where the interpolation points coincide with a
suitable quadrature for the mass matrix. The case of non affine
mapping is especially relevant for cube grids and here tensor product
quadrature points can be used to construct the Lagrange type space. A
possible approach is to use a tensor product Gaussian rule which
results in a diagonal mass matrix for each element in the grid with a
trivial dependency on the element geometry. This is due to the fact that
this quadrature is accurate up to order $2k+1$ so that it
can be used to exactly compute the mass matrix
$\int \varphi_i\varphi_j = \sum_q \omega_q \varphi_i(x_q)\varphi_j(x_q)
   = \sum_q \omega_q\delta_{iq}\delta_{jq}=\omega_i\delta_{ij}$ 
which will therefore be diagonal with diagonal elements only depending on the integration
element at the quadrature points. In addition, if
all the other element integrals needed to evaluate the spatial operator use
the same quadrature, the interpolation property of the basis functions can be used
to further speedup evaluation. All of this is well known and for example
investigated in \cite{kopriva:02}. 
A second common practise is to use the points of a tensor product
Lobatto-Gauss-Legendre (LGL) quadrature (see e.g. \cite{kopriva:10}).
In this case the evaluation of the intersection integrals in the spatial
operator can also be implemented more efficiently. To still retain the
simple structure of the mass matrix requires to use the same
Lobatto-Gauss-Legendre rule to compute $\int \varphi_i\varphi_j$. Since
this rule is only accurate up to order $2k-1$ this results in an
underintegration of the mass matrix similar to mass lumping;
this does not seem to influence the accuracy of the scheme.
This approach is often referred to a spectral discontinuous Galerkin method
\cite{kopriva:02,kopriva:10}.

After fixing the grid and the discrete space,
we seek $\df \in \phispace$ by discretizing the spatial operator
$\oper{L}(\sol)$  in \eqref{eqn:general_op}
with either Dirichlet, Neumann, or Robin type boundary conditions
by defining for all test functions $\basefct \in \phispace$, 
\begin{equation}
\label{convDiscr}
\dual { \spcoper(\df) } := \dual{ K_h(\df) } + \dual{ {I}_h(\df) }
\end{equation}
with the element integrals 
\begin{eqnarray}
\label{eqn:elementint}
   \dual{ {K}_h(\df) } &:=&
      \sum_{\elem \in \grid} \int_{\elem}
      \big( ( \Fc(\df) - \Fv(\df, \nabla \df ) ) : \nabla\basefct + \su
      \cdot \basefct \big),
\end{eqnarray}
with $\su = \Simpl(\df) + \Sexpl(\df)$ and the surface integrals (by introducing appropriate numerical fluxes 
$\fluxF$, $\fluxA$ for the convection and diffusion terms, respectively) 
\begin{eqnarray}
\label{eqn:surfaceint}
   \dual{ {I}_h(\df) } &:=&
      \sum_{e \in \Gamma_i} \int_e \big(
      \vaver{\Fv(\df, \vjump{\df} )^T : \nabla\basefct} +
      \vaver{\Fv(\df, \nabla\df)} : \vjump{\basefct} \big) \nonumber \\
    &&- \sum_{e \in \Gamma} \int_e \big( \fluxF(\df) - \fluxA(\df,\nabla\df)\big) :
      \vjump{\basefct},
\end{eqnarray}
with $\vaver{ \sol }, \vjump{ \sol }$ denoting the classic average and jump of $\sol$ over $e$,
respectively.

The convective numerical flux $\fluxF$ can be any appropriate numerical flux known for
standard finite volume methods.
e.g. $\fluxF$
could be simply the local Lax-Friedrichs flux function (also known as Rusanov
flux)
\begin{equation}
  \label{flux:llf}
  \fluxF^{LLF}(\df)|_{\isec} :=
      \vaver{ \Fc(\df) } + \frac{\lambda_{\isec}}{2}
         \vjump{ \df } 
\end{equation}
where $\lambda_{\isec}$ is an estimate of the maximum wave speed on intersection
$\isec$. 
One could also choose a more problem tailored flux (i.e. approximate Riemann solvers).
Different options are implemented in \dunefemdg (cf. \cite{dunefemdg:17}).

A wide range of diffusion fluxes $\fluxA$ can be found in the
literature and many of these fluxes are available in \dunefemdg, for example, 
Interior Penalty and variants, Local DG, Compact DG 1 and 2 
as well as Bassi-Rebay 1 and 2 (cf. \cite{cdg2:12,dunefemdg:17}).

\subsection{Temporal discretization}
\label{TimeDisc}

To solve the time dependent problem~\eqref{eqn:general} we use a method of
lines approach in which the DG method described above is first used to
discretize the spatial operator and then a solver for ordinary differential
equations is used for the time discretization, see \cite{dunefemdg:17}.
After spatial
discretization, the discrete solution $\df(t) \in \phispace$ 
has the form $\df(t,x) = \sum_i \sol_i(t)\basefct_i(x)$.
We get a system of ODEs for the coefficients of $\sol(t)$ which reads 
\begin{eqnarray}
  \label{eqn:ode}
  \sol'(t) &=& f(\sol(t))  \mbox{ in } (0,T]
\end{eqnarray}
with $f(\sol(t)) = M^{-1}\spcoper(\df(t))$, $M$ being the mass matrix which is in
our case is block diagonal or even the identity, depending on the choice of basis
functions. $\sol(0)$ is given by the projection of $\sol_0$ onto $\phispace$.
%
For the type of problems considered here the most popular choices for the
time discretization are based around the ideas of
\textit{Strong Stability Preserving} Runge-Kutta methods (SSP-RK) (for details see
\cite{dunefemdg:17}). Depending on the problem SSP-RK method are available
for treating the ODE either explicitly, implicitly, or by additively splitting the
right hand an implicit/explicit (IMEX) treatment is possible. In an example
for such a splitting
$\oper{L}(\sol) = \oper{L}_{e}(\sol) + \oper{L}_{i}(\sol)$ 
the advection term would be treated explicitly while the diffusion term
would be treated implicitly. In addition the source term could be split as
well leading to operators of the form
\begin{eqnarray}
  \oper{L}_{e}(\sol) := - \nabla \cdot \Fc(\sol) + \Sexpl(\sol) &\quad \mbox{and}\quad &
  \oper{L}_{i}(\sol) := \nabla \cdot \Fv(\sol,\nabla\sol) \big) + \Simpl(\sol) 
\end{eqnarray}
Of course other splittings are possible but currently not
available through the Python bindings.

We implement the implicit and semi-implicit Runge-Kutta solvers using a Jacobian-free Newton-Krylov
method (see \cite{knoll:04}).

\subsection{Stabilization}
\label{Stabilization}

The RK-DG method is stable when applied to linear problems such as linear hyperbolic systems; 
however for nonlinear problems spurious oscillations occur near strong shocks or steep gradients.
In this case the RK-DG method requires some extra stabilization. 
In fact, it is well known that only the first order scheme ($k=0$) 
produces a monotonic structure in the shock region.
Many approaches have been suggested to make this property available in 
higher order schemes, without introducing the amount of
numerical viscosity, which is such a characteristic feature of first order schemes.
Several approaches exist, among those are slope limiters \cite{CSreview,rivlimiter:06,limiter:11,Krivodonova2004,Dumbser:08} 
and for a comprehensive literature list we refer to \cite{Shu-boundpreserving-review:16}.
Another popular stabilization technique is the artificial diffusion (viscosity) 
approach \cite{feist:08,Kloeckner,persson:06,Guermond:11,Discacciati:20} and
others. Further techniques exists, such as a posteriori techniques to stabilize the DG method
\cite{DMO:07} or order reduction \cite{Dolejsi:03}. 

In the following, we focus on an approach based on limiting combined with a troubled cell
indicator. We briefly recall the main steps since these are necessary to understand the code design decision later on. 
Note that an artificial diffusion approach could be used easily by adding a
suitable diffusion operator to the model described above.

A stabilized discrete operator
is constructed by concatenation of the DG operator $\spcoper$ from \eqref{convDiscr} and a stabilization 
operator $\limiter$, leading to a modified discrete spatial operator $\limitedoper(\df) := (\spcoper \circ
\limiter)(\df)$. 
The stabilizations considered in this work can 
be computed element wise based only on data from neighboring elements thus
not increasing the stencil of the operator.
Given a DG function $\df$ we call $\df^* = \limiter(\df)$ the stabilized DG function
and we call $\dfelem = \df|_{\elem}$ the restriction of a function $\df$ on element $\elem$
and denote with $\ubar_\elem$ its average. 
Furthermore, we call $\isec$ the intersection 
between two elements $\elem,\neig$ for $\neig \in \neigh_{\elem}$, with $\neigh_{\elem}$ being the 
set of neighbors of $\elem$ and $\interset$ the set of intersections of $\elem$
with it's neighbors.


The stabilized solution should fulfil the following requirements: 
\begin{enumerate} 
  \item \textit{Conservation property}, i.e. $\ubar_{\elem} = \ubar^*_{\elem}$
  \item 
    \label{req:physicallity}
    \textit{Physicality of $\sol^*_{\elem}$} (i.e. values of $\sol^*_{\elem}$ belong to the set of
    states, i.e. positive density etc.) at least for all quadrature points used to
    compute element and surface integrals. 
  \item 
    \label{req:troubledcell}
    \textit{Identity in "smooth" regions}, i.e. in regions where the solution 
    is "smooth" we have $\df^* = \df$. 
    This requires an indicator for the smoothness of the solution. 
  \item \textit{Consistency for linear functions}, i.e. 
    if the average values of $\df$ on $\elem$ and its neighbors are given by the same linear function $L_{\elem}$ 
    then $\sol^*_{\elem} = L_{\elem}$ on $\elem$.
  \item 
    \label{req:minimalstencil}
    \textit{Minimal stencil}, i.e. the stabilized DG operator shall have the same stencil as 
    the original DG operator (only direct neighboring information in this
    context).
  \item \textit{Maximum-minimum principle and monotonicity}, i.e. in regions where the solution 
    is not smooth the function $\sol^*_{\elem}$ should only take values between 
     $\min_{\neig \in \neigh_{\elem}} \ubarneig$ and 
     $\max_{\neig \in \neigh_{\elem}} \ubarneig$.
\end{enumerate}

These requirements are built into the stabilization operator in different ways. 
Requirement \ref{req:minimalstencil}
simply limits the choice of available reconstruction methods, e.g. higher order
finite volume reconstructions could not be used because of the potentially
larger stencil that would be needed which is not desired here. 
For example, requirement \ref{req:physicallity} and \ref{req:troubledcell} 
form the so called \textit{troubled cell indicator} that triggers whether a
stabilized solution has to be computed.
The stabilization then consists of two steps 
\begin{enumerate} 
  \item We define the set of \textit{troubled cells} by  
    \begin{align}
        TC(\df) := \{ \elem \in \grid \, \colon \, \indicator_{\elem}(\df) > TOL 
                      \text{ or $\dfelem$ has \textbf{unphysical values}} \}
      \label{eq:troubledcell}
    \end{align}
       with $\indicator_{\elem}$ being a smoothness indicator detailed in \eqref{shockIndi} and $TOL$ a threshold that can be influenced by the user.
   \item
     \begin{enumerate}
       \item If $\elem \notin TC(\df)$ then $\dfelem^*  = \dfelem$, i.e.
       the operator $\Pi_h$ is just the identity. 

      \item \textit{Construction of an admissible DG function}
       if $\elem\in TC(\df)$.
       In this case the DG solution on $\elem$ needs to be altered until $\elem$ is no longer a troubled cell.
       In this work we guarantee this by
       restricting ourselves to the reconstruction of limited linear functions based on
       the average values of $\df$ on $\elem$ and it's neighbors, similar
       to 2nd order MUSCL type finite volume schemes when used with
       piecewise constant basis functions.
    \end{enumerate}
\end{enumerate}

\subsection{Adaptivity}
\label{sec:dgAdaptivity}

Due to the regions of steep gradients and low regularity in the solution, advection dominated
problems benefit a lot from the use of local grid refinement and coarsening.
In the example shown here we use a residual type estimator where the
adaptation process is based on the residual of an auxiliary PDE
\begin{equation}
  \partial_t \eta(t,x,U) + \nabla \cdot ( F(t,x,U,\nabla U) )  =  S(t,x,U,\nabla U)
\end{equation}
where for example $\eta,F,S$ could simply be taken from one of the
components of the PDE or could be based on an entropy, entropy flux pair.
We now define the element residual to be
\begin{align}
\label{eq:estimator}
\eta_{E}^{2}=& h^{2}_{E}\norm{R_{vol}}^{2}_{L^{2}(E)}+\frac{1}{2}\sum\limits_{e \in \interset}\left( \ h_e \norm{R_{e_2}}^{2}_{
L^{2}(e)}+\frac{1}{h_e}\norm{R_{e_1}}^{2}_{L^{2}(e)} \ \right) 
\end{align} 
where $R_{vol}$ is a discretized version of the interior residual indicating how accurate the
discretized solution satisfies the auxiliary PDE at every interior point of the domain for two timesteps $t^n$ and $t^{n+1}$,
\begin{align*}
  R_{vol} := & \frac{1}{t^{n+1}-t^n}\big( \eta(t^{n+1},\cdot,U^{n+1})-\eta(t^n,\cdot,U^n) \big) \\
  & \hspace*{1cm}
         + \frac{1}{2} \nabla \cdot \big ( F(t^n,\cdot,U^n) + F(t^{n+1},\cdot,U^{n+1}) \big )
         - \frac{1}{2} \big (S(t^n,\cdot,U^n) + S(t^{n+1},\cdot,U^{n+1}) \big ) .
\end{align*}
and we have jump indicators
\begin{align*}
  R_{e_2}   := \frac{1}{2} \vjump{ F(t^n,\cdots) + F(t^{n+1},\cdots) } & \qquad
  R_{e_{1}} := \frac{1}{2} \vjump{ \eta(t^n,\cdots) + \eta(t^{n+1},\cdots) }.
\end{align*}
This indicator is just one of many possible choices, e.g., simply taking the jump
of $\df$ over intersections will often lead to very good results as well.
We have had very good experiences with this indicator as shown here based on 
earlier work (see \cite{twophase:18}) where we used a similar indicator. 
Related work in this direction can be found in \cite{DMO:07}
where a similar indicator was derived.


\subsection{Summary of building blocks and limitations}

\begin{itemize}
  \item \textbf{Grid structure:}
      As discussed in Section~\ref{sec:dune} DG method have very few
      restrictions on the underlying grid structure. Although more general
      method have been implemented based on \dunefemdg we restrict our
      attention here on method that can be implemented using direct neighboring information only. 
      This excludes for example direct implementation of the \emph{local DG
      method} and also more general reconstruction procedures. From a
      parallelization point of view, the use of a minimal stencil is beneficial
      for many core architectures, but crucially available unstructured grids in \dune only 
      implement this ghost cell approach and arbitrary overlap is only available for Cartesian grids. 
      In the next section we will show results for Cartesian, general cube
      and simplex grids, and also some results using polyhedral elements.
  \item \textbf{Space:}
      DG methods are variational methods based on discrete function spaces
      defined over the given grid. As mentioned in Section \ref{seq:discretization_spatial}
      the efficiency of an implementation will depend not only on the space
      used but also on the choice of basis functions. We will show results
      for a number of different choices in the following section.
   \item \textbf{Numerical fluxes:} the discretization of both the diffusion and
      advective terms, involve boundary integrals where the discrete
      solution is not uniquely defined requiring the use of numerical fluxes.
      As mentioned, we only consider methods with a minimal stencil so that
      fluxes have to be used which depend only on the traces of the discrete solution
      on both sides of the interface. In the package
      presented here a large number of fluxes for the diffusion are
      available (including but not limited to (non-symmetric) interior penalty,
      Bauman-Oden, Bassy-Rebay 1 and 2, Compact DG 1 and 2).
      Since we are focusing on advection dominated problems we will not
      discuss this aspect further. For the advective numerical flux, the simplest choice for the
      advective flux is the Local-Lax-Friedrichs or Rusanov flux from equation
      \eqref{flux:llf} which requires little additional input from the user and in combination
      with higher order schemes generally gives good results. For the Euler
      equations of gas dynamics a number of additional fluxes are available
      in \dunefemdg, including well known fluxes like HLL, HLLC, and HLLEM.
      Since using a problem specific flux for the hyperbolic flux can be a
      good way to guarantee that the method has some additional properties,
      e.g., well balancing for shallow water flow, we will shortly touch on
      the possibility of implementing new numerical fluxes in the following
      section.
    \item \textbf{Stabilization:} as was discussed in some detail in the
      previous section, our stabilization approach is based on
      a troubled cell indicator (combining a smoothness indicator with a
      check on physicallity at quadrature points) and a reconstruction
      process of the solution in troubled cells. While at the time of
      writing it is not straightforward to implement a new reconstruction
      strategy without detailed knowledge of the C++ code, it is fairly
      easy for a Python user to provide their own troubled cell indicator
      which is demonstrated in the next section.
    \item \textbf{Time evolution:} after the spatial discretization, the resulting
      ordinary differential equation is solved using a Runge-Kutta as
      already mentioned. \dunefemdg provides a number of explicit,
      implicit, and IMEX SSP-RK method up to order four. But since this part of the algorithm
      is not so time critical compared to the evaluating the discrete
      spatial operator we will also show in the following section how a
      different time evolution method can be implemented on the Python side.
\end{itemize}

In the following we will describe how the different building blocks making
up the DG method can be provided by the user, starting with some very
simple problems, requiring little input from the user and slowly building
up the complexity. We are not able to provide the full code listings in
each case but a tutorial with all the test cases presented here is
available as discussed in Appendix~\ref{sec:installation}
and also available from \cite{dunefemdgrepo}.

\section{Customizing the DG method using the Python Interface}
\label{sec:implementation}
In the following we describe how to use the DG methods provided by
\dunefemdg based on the mathematical description provided in the previous
section. The setup of a simulation always follows the same structure which
is similar to the discussion in Section~\ref{sec:equations}. The corresponding Python code is
discussed in Section~\ref{sec:simsetup}.
First a \pyth{gridView} is constructed given a
description of (coarse) tessellation of the computational domain. The same
basic description can often be used to define grids with different
properties, i.e., different element types or refinement strategies.
This \pyth{gridView} is then used to construct a discrete space - again
different choices are available as discussed in
Section~\ref{seq:discretization_spatial}. The space is then used to
construct the discrete version of the spatial operator $\oper{L}$
in (\ref{eqn:general}). Following the method of lines approach this discrete
operator is then used to construct an ODE solver. The section concludes with the basic time loop
which evolves the solution from $t=0$ to $t=T$.
If not stated otherwise, all the examples in this section are based on the code described in
Section~\ref{sec:simsetup}. The details of the PDE we want to solve are
encoded in a \pyth{Model} class. This contains only information about the
continuous problem, i.e., not about the discretization details. All
attributes that can be set on this class are summarized in Appendix~\ref{sec:extension}.
Its main purpose is to define the boundary conditions, fluxes, and source term functions
defining the spatial operator $\oper{L}$ in (\ref{eqn:general}).
It is then used to construct the discrete spatial operator.
For a scalar first order hyperbolic problem for example
this requires the definition of $\Fc$, shown here for a simple linear
advection problem:
\begin{python}[Basic Model class for advection problem with constant velocity $(1,2)$]
class Model:
    dimRange = 1      # scalar problem
    def F_c(t,x,U):   # advective flux
        return as_matrix( [[ U[0],2*U[0] ]] )
\end{python}
We add additional information like the description of the computational
domain, the final time of simulation $T$ and the initial conditions not
strictly required for the spatial operator. This simplifies the description
of the setup process in the next section. In our fist example 
in Section~\ref{sec:linearAdv} we start with a similar advection problem as
the one given above and then step by step how the \pyth{Model} class is
extended to include additional features, like the stabilization, and
adaptivity. In Section~\ref{sec:euler} we show how a vector valued PDE is
implemented using the example of the Euler equations of gas dynamics. In
this section we also include an example showing how a source term is added
and also how a diffusive flux is added by providing an implementation of
the compressible Navier-Stokes equations.

In the next two sections we demonstrate how to extend the existing code.
In Section~\ref{sec:userdeftroubledcell} we discuss how to implement a
different troubled cell indicator and in Section~\ref{sec:userstepper} how
a user defined time stepper can be used.

We conclude this section with a discussion on the use of different spaces
and grid structures (Section~\ref{sec:gridspace}) and show a
simulation for an advection-diffusion-reaction problem in
Section~\ref{sec:radproblem}.

If not mentioned otherwise, the results presented in the following are
obtained using a Cartesian grid with a forth order polynomial space spanned
by orthonormal basis function. We use the local Lax Friedrichs flux \eqref{flux:llf} for the advection
term and the CDG2 method \cite{cdg2:12} for the problems with diffusion. The troubled cell
indicator is based on a smoothness indicator discussed in \cite{limiter:11}.
The reconstruction is computed based on a linear programming problem as described in
\cite{Chen:LPreconstruction}. Finally, we use a standard third order SSP Runge-Kutta method for the time evolution
\cite{shu:01}.

\subsection{Simulation setup}
\label{sec:simsetup}
First we need to choose the grid structure and the computational domain.
For simplicity we concentrate here on problems defined on
2D intervals $[x_l,x_r]\times [y_b,y_t]$ and use an axis aligned cube grid
capable of non-conforming refinement, with $n_x,n_y$ elements in $x$ and $y$ direction, respectively.
The space consists of piecewise polynomials of degree $p=4$ spanned by an
orthonormal basis over the reference element $[0,1]^2$. This can be setup
with a few lines of Python code:
\begin{python}[Setting up the grid and discrete function space]
from dune.alugrid import aluCubeGrid as grid
from dune.fem.view import adaptiveLeafGridView as view  # needed for adaptive simulation
from dune.fem.space import dgonb

gridView = view( grid( Model.domain ) )
gridView.hierarchicalGrid.globalRefine(3)   # refine a possibly coarse initial grid
space    = dgonb( gridView, dimRange=Model.dimRange, order=4 ) # degree p = 4 
U_h      = space.interpolate(Model.U0, name="solution")
\end{python}
As already mention, the \pyth{Model} class combines information on the PDE and as seen above also provides
a description of the domain, the number of components of the PDE to solve and the initial
conditions $U_0$ defined as a UFL expression. Examples will be provided in
the next sections, an overview of all attributes on this class is given in Appendix~\ref{sec:extension}.
The discrete function \pyth{U_h} will contain our solution at the current
time level and is initialized by interpolating $U_0$ into the discrete
function space.

After setting up the grid, the space, and the discrete
solution, we define the DG operator and the default Runge-Kutta time stepper:
\begin{python}[Setting up the spatial operator and the time evolution scheme]
from dune.femdg import femDGOperator
from dune.femdg.rk import femdgStepper

operator = femDGOperator(Model, space, limiter=None)
stepper  = femdgStepper(order=3, operator=operator) # consistency order 3 
\end{python}
\label{opcode}
The \pyth{Model} class is used to provide all required information to
the DG operator. In addition, the constructor of the DG operator takes a number of
parameters which will be described throughout this section as required
(see also Appendix~\ref{sec:extension}).
In our first example we will require no stabilization so the \pyth{limiter}
argument is set to \pyth{None}.
In the final line of code, we pass the constructed operator to the time stepper together with
the desired order (for the following simulations we use methods of consistency order $3$).
Finally, a simple loop is used to evolve the solution from the starting
time (assumed to be $0$) to the final time $T$ (which is again a property of the \pyth{Model} class):
\begin{python}[Time loop]
t = 0
while t < Model.endTime:
    dt = stepper(U_h)
    t += dt
\end{python}
The call method on the stepper evolves the solution $U_h$ from the current
to the next time step and returns the size $\Delta t$ for the next
time step.
In the following we will describe the \pyth{Model} class in
detail and show how the above code snippets have to be modified to include
additional features like stabilization for nonlinear advection problems or adaptivity.
We will not describe each problem in detail, the provided code should give
all the required information.


\subsection{Linear advection problem}
\label{sec:linearAdv}

\begin{figure}
\centering
  \subfloat[none]{
    \includegraphics[width=0.48\textwidth]{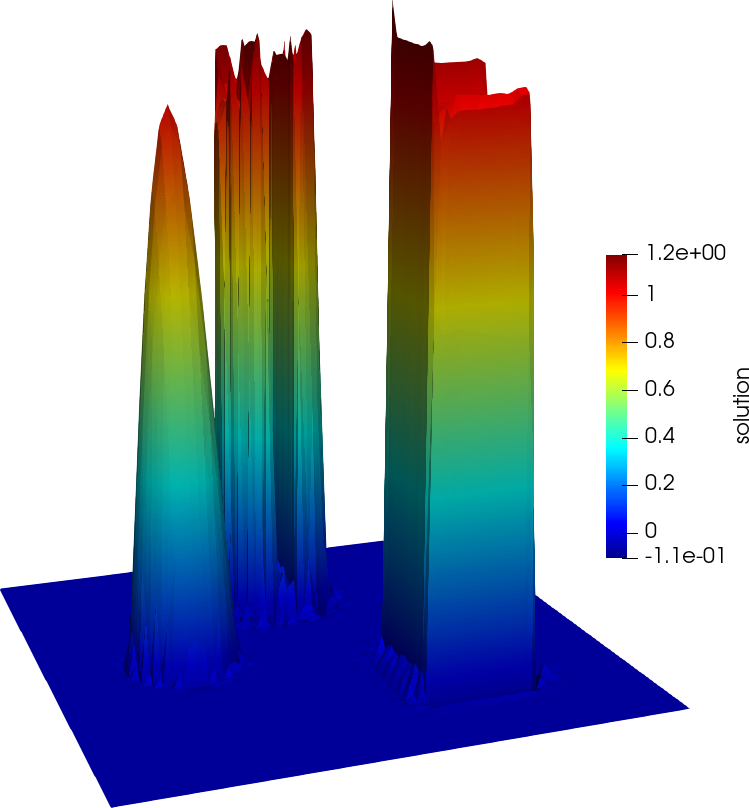}}
  \subfloat[default stabilization]{
    \includegraphics[width=0.48\textwidth]{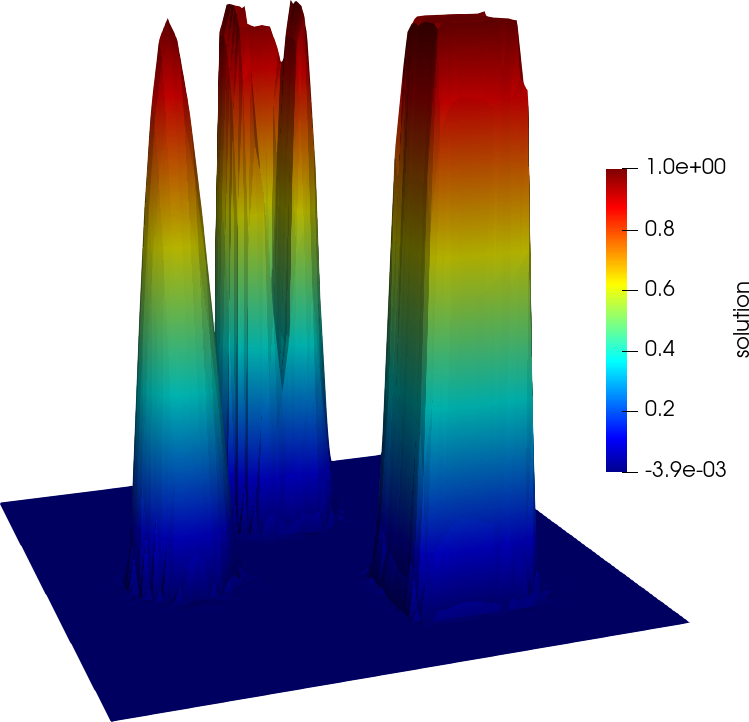}}
  \caption{Three body advection problem at time $t=\pi$. (a) without
  stabilization and (b) with limiter based stabilization.
Color range based on minimum and maximum values of discrete solution.}
\label{fig:advectionnone}
\end{figure}

In the following we solve a linear advection problem and explain 
various options for stabilization and local grid adaptivity.
The advection velocity is given by $(-y,x)$ so that the initial conditions
are rotated around the origin. The initial conditions
consist of three parts with different regularity:
a cone, the characteristic function of a square and a slotted circle.
Using UFL these can be easily defined:
\begin{python}[Velocity and initial conditions for three body advection problem]
from dune.ufl import cell
from dune.grid import cartesianDomain
from ufl import *

x = SpatialCoordinate( cell(2) ) # setup a 2d problem
y = x - as_vector([ 0.5,0.5 ])
velocity = as_vector([-y[1],y[0]])      # velocity

# setup for 'three body problem'
# body 1: cube
cube  = ( conditional( x[0]>0.6, 1., 0. ) * conditional( x[0]<0.8, 1., 0. )*
          conditional( x[1]>0.2, 1., 0. ) * conditional( x[1]<0.4, 1., 0. ) )
# body 2: slotted cylinder with center (0.5,0.75) and radius 0.1
cyl = ( conditional( (x[0] - 0.5)**2 + (x[1] - 0.75)**2 < 0.01, 1.0, 0.0 ) *
       ( conditional( abs(x[0] - 0.5) >= 0.02, 1.0, 0.0)*
         conditional( x[1] < 0.8, 1.0, 0.0) + conditional( x[1] >= 0.8, 1., 0.) ) )
# body 3: smooth hump with center (0.25,0.5)
hump = ( conditional( (x[0] - 0.25)**2 + (x[1] - 0.5)**2 < 0.01, 1.0, 0.0 ) *
         2/3*(0.5 + cos(pi * sqrt( (x[0] - 0.25)**2 +  (x[1] - 0.5)**2 ) / 0.15 )) )
\end{python}
We will solve the problem on the time interval $[0,\pi]$ resulting in half
of a rotation.

\subsubsection{Model description}
\begin{python}[Basic Model class for three body advection problem]
class Model:
    dimRange = 1
    endTime = pi
    U0      = [cube+cyl+hump]
    domain  = cartesianDomain((0,0),(1,1),(10,10))
    def F_c(t,x,U):
        return as_matrix( [[ *(velocity*U[0]) ]] )
    def maxWaveSpeed(t,x,U,n):
        return abs(dot(velocity,n))
    # simple 'dirchlet' boundary conditions on all boundaries
    boundary = {range(1,5): lambda t,x,U: as_vector([0])}
\end{python}
Note that to use the local Lax-Friedrichs flux from \eqref{flux:llf} 
we need to define the analytical flux function \pyth{F_c(t,x,U)} and 
the maximum wave speed \pyth{maxWaveSpeed(t,x,U,n)} in the Model class of the hyperbolic problem. This function is also used to provide an
estimate for the time step based on the CFL condition, which is returned by the \pyth{stepper}. The
stepper also provides a property \pyth{deltaT} which allows to fix
a time step to use for the evolution.
A result for the described \textit{three body problem} is presented in Fig. \ref{fig:advectionnone} (left)
and Fig. \ref{fig:advectionminmod} (left).

\subsubsection{Stabilization}
It is well known that the DG method with a suitable numerical flux is
stable when applied to linear advection problems like the one studied here.
But it does produce localized over and undershoots around steep gradients.
While this is not necessarily a
problem for linear equations, it can be problematic in some instances,
e.g., when negative values are not acceptable. More crucially this
behavior leads to instabilities in the case of nonlinear problems.
Therefore, a stabilization mechanism has to be provided. As mentioned in
Section~\ref{Stabilization}, we use a reconstruction approach combined with a troubled cell indicator.

The implementation available in \dunefemdg is based on a troubled cell
indicator using a smoothness indicator suggested in \cite{Krivodonova2004}:
this indicator accumulates the integral of the jump of some scalar
quantity derived from $\df$ denoted with $\phi(\dfelem, \dfneig)$ over
all inflow boundaries of $\elem$
(defined as the intersection where some given velocity $\vecv$ and
the normal $\ne$ have opposite signs), i.e.
\begin{align}
\label{shockIndi}
\indicator_{\elem}(\df) := \sum_{\suminter{\interset}{<} }
                      \left (
                       \frac{\int_{\isec} \phi(\dfelem, \dfneig) \, ds}
                        {\alpha_d(k) \, h^{(k+1)/4}_{\elem}\, |\isec| }
                        \right ) ,
\end{align}
$\alpha_d(k) = \frac{2}{125}d 5^k$ denotes a scaling factor,
$h_{\elem}$ is the elements diameter and $|\isec|$
the area of the intersection between the two elements $\elem$ and
$\neig$. Please note the slight derivation from the notation in 
\cite{limiter:11} by denoting the smoothness indicator 
in \eqref{shockIndi} with $\indicator_{\elem}$ instead of $S_{\elem}$. 

A cell $\elem$ is then flags as \emph{troubled} if
$\indicator_{\elem}(\df) > TOL$
where $TOL$ si a threshold that can be set by the user.
As default we use $TOL=1$. On such \emph{troubled} cells a reconstruction
of the solution is used to obtain a suitable approximation.
Two reconstruction methods are implemented, one described in \cite{limiter:11} and extended to general
polyhedral cells in \cite{reconpoly:17} and an optimization based
strategy described in \cite{Chen:LPreconstruction} based on ideas from
\cite{May2013}. Both strategies correspond to 2nd order MUSCL type finite volume schemes when used with
piecewise constant basis functions.

\begin{figure}
\centering
  \subfloat[none]{
    \includegraphics[width=0.48\textwidth]{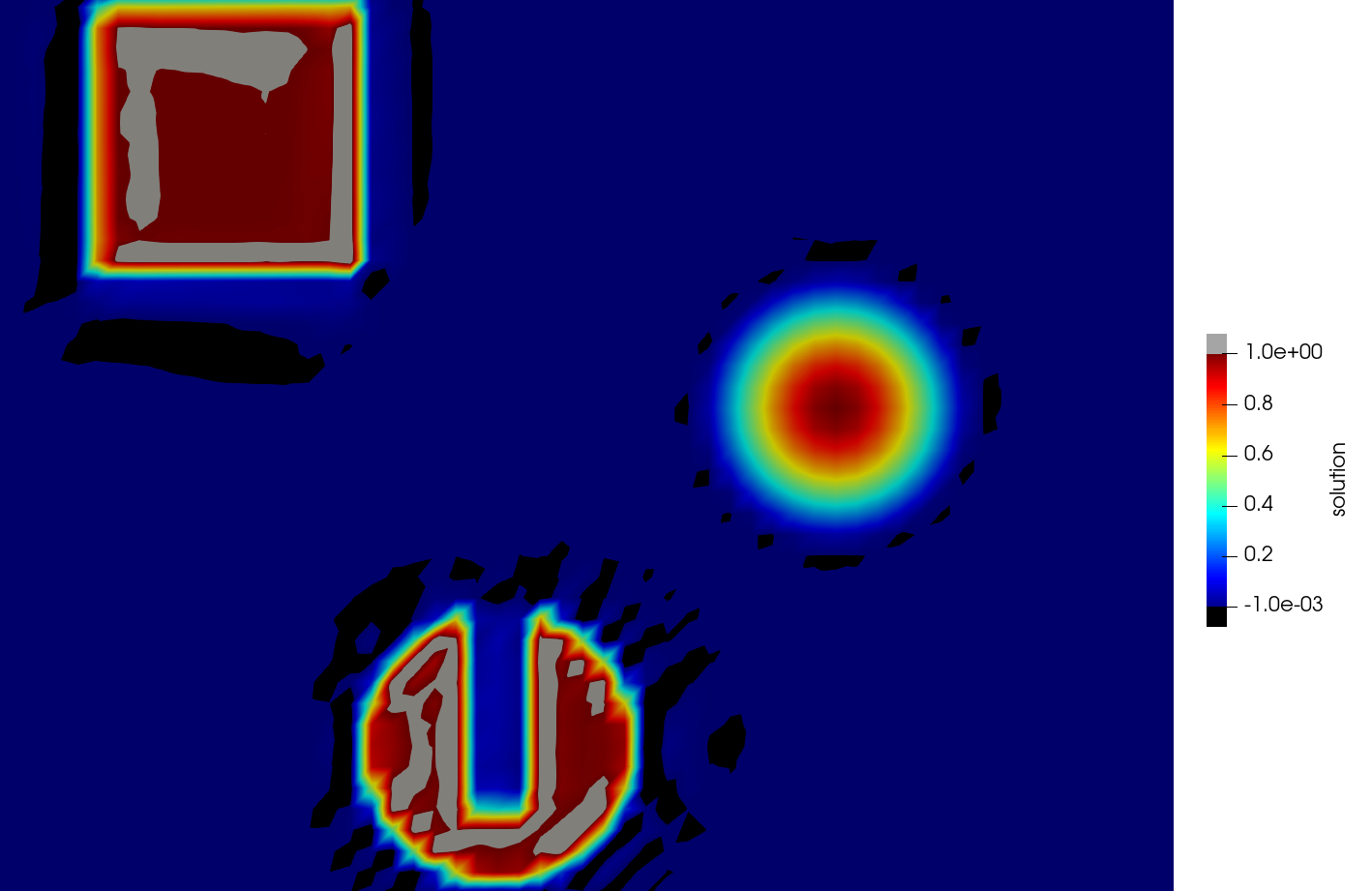}}
  \subfloat[default stabilization]{
    \includegraphics[width=0.48\textwidth]{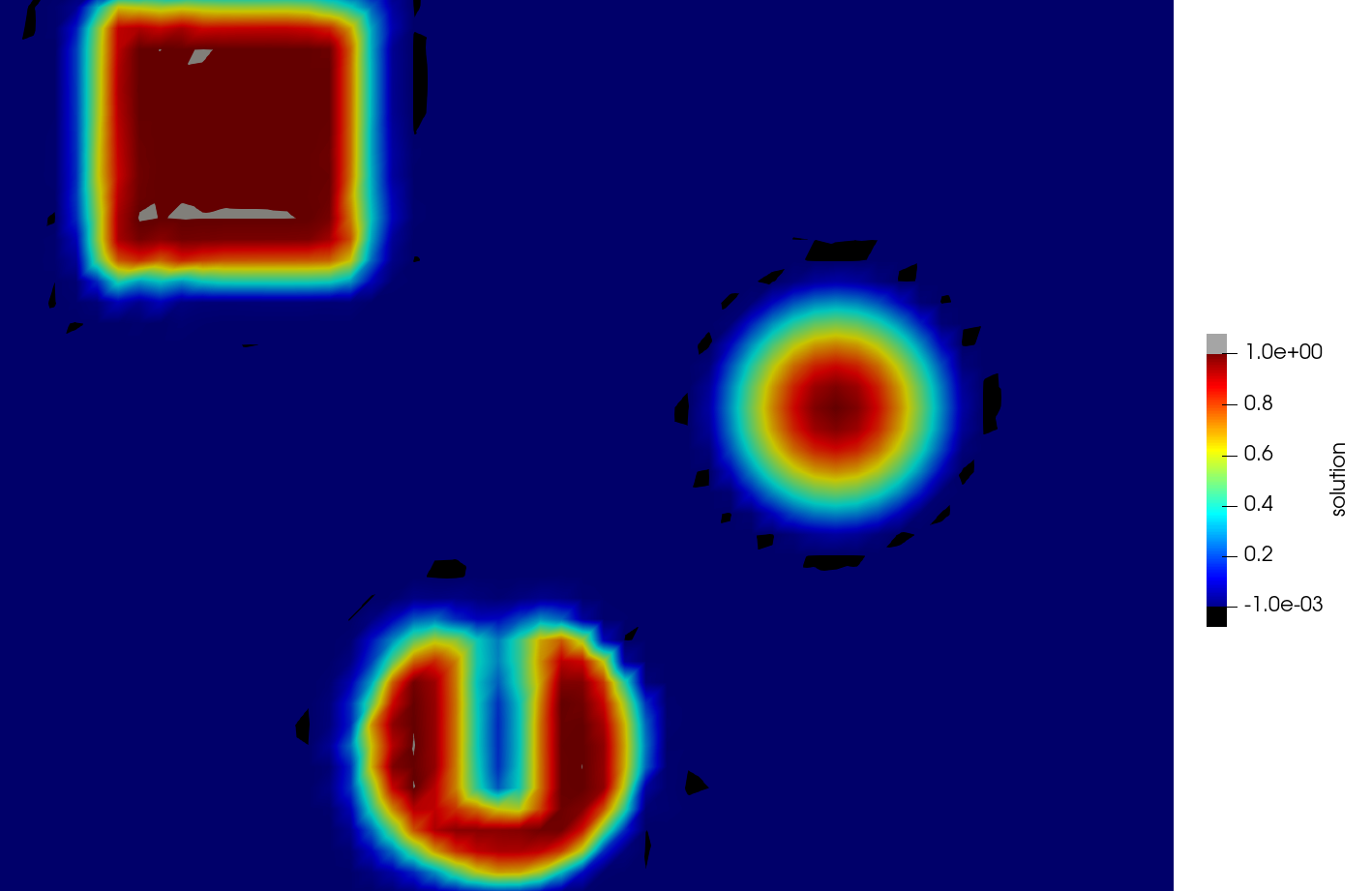}} \\
  \subfloat[default \& physical]{\includegraphics[width=0.48\textwidth]{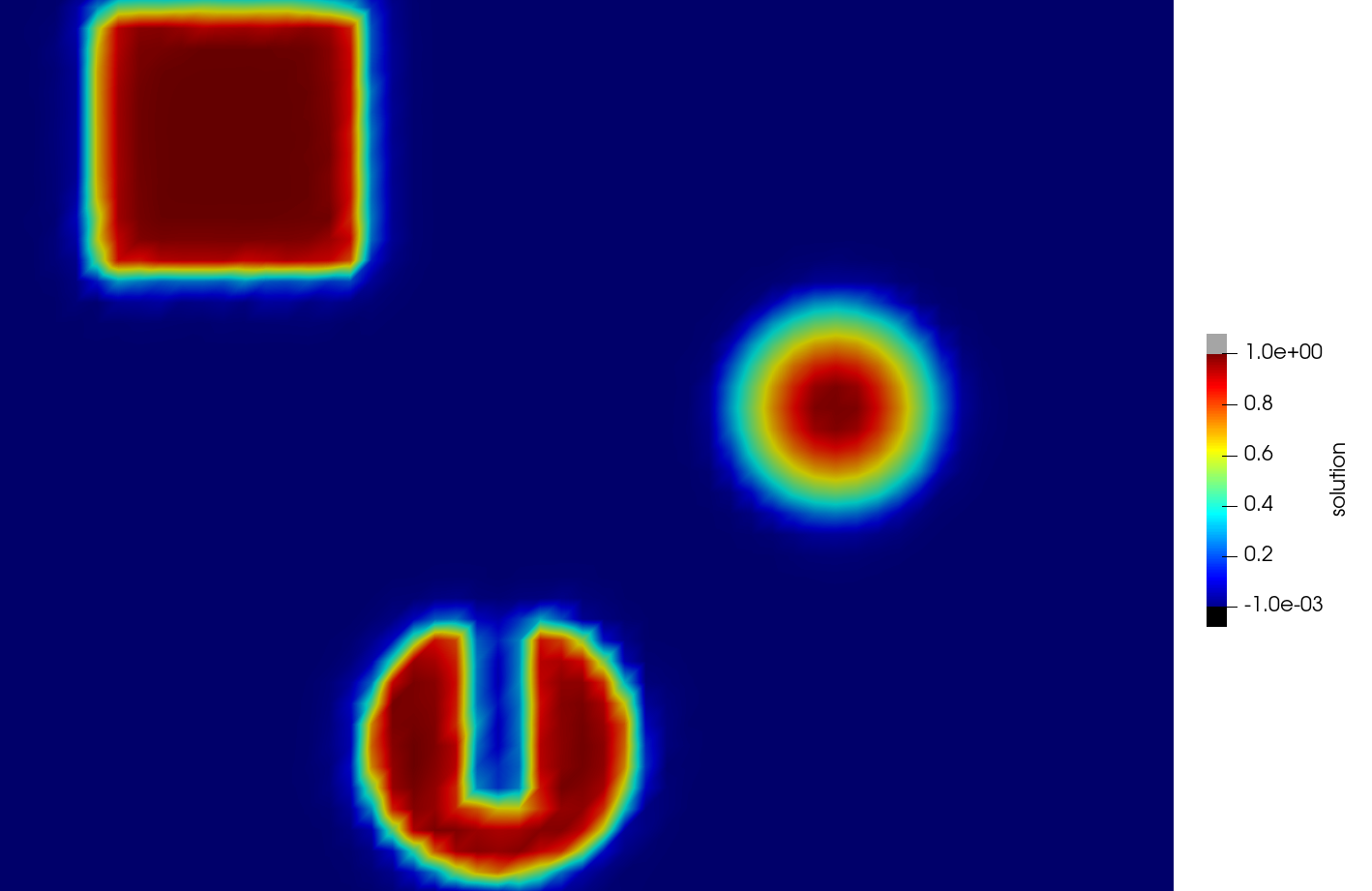}}
  \subfloat[scaling]{\includegraphics[width=0.48\textwidth]{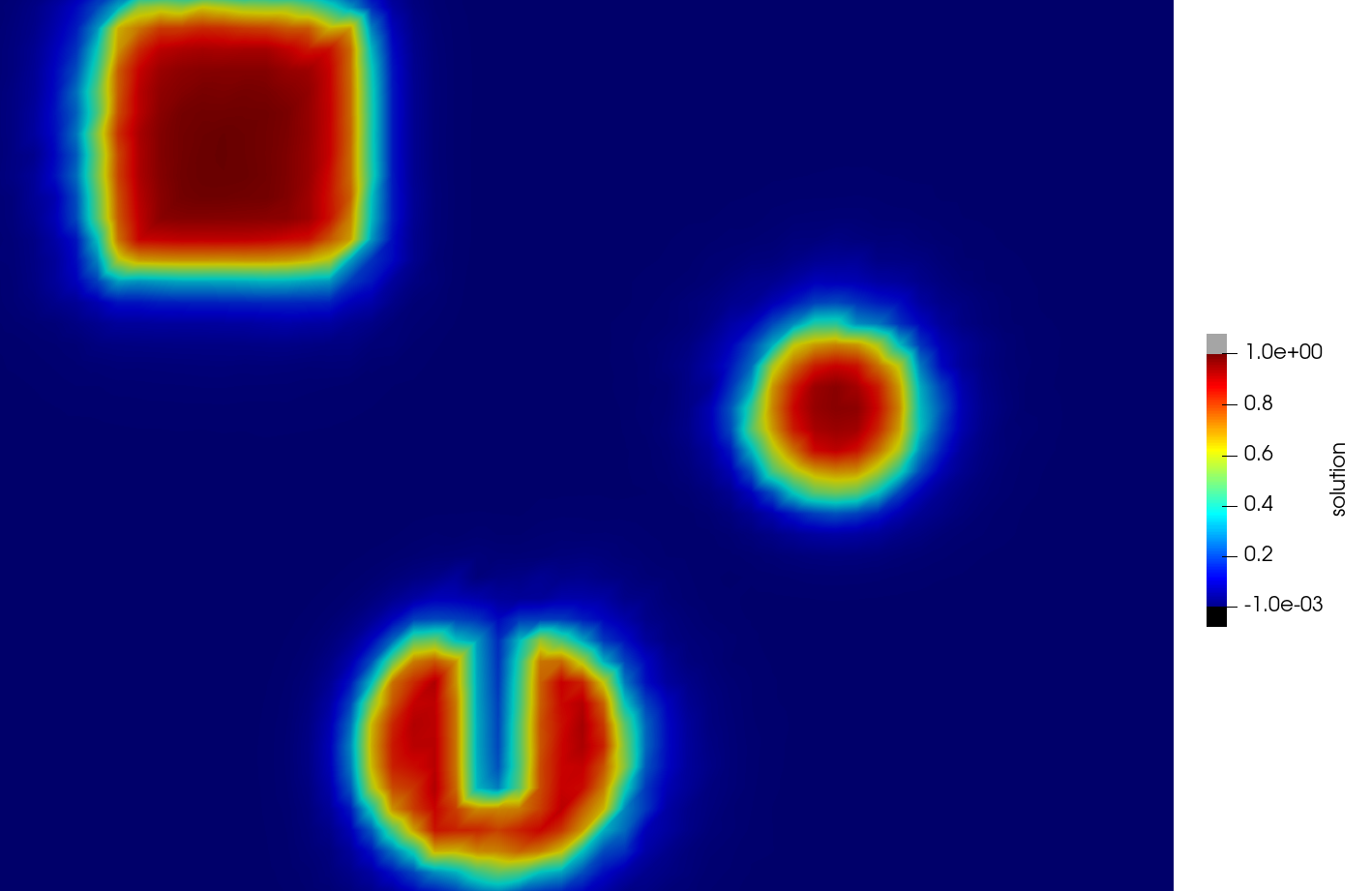}}
\caption{Top view of three body advection problem at time $t=\pi$.
  (a) without stabilization, (b) with default limiter based stabilization using, 
  (c) using physicality check on top of the default limiter to reduce oscillations, and (d) using a scaling limiter.
Values above $1+10^{-3}$ or below $-10^{-3}$ indicate oscillations and are colored in grey and black, 
  respectively. Without stabilization over and undershoots were around $20\%$ (see
  Figure \ref{fig:advectionnone}) and completely removed by adding the
  \emph{physicality} check.}
\label{fig:advectionminmod}
\label{fig:advection_phys_scaling}
\end{figure}

Therefore, the stabilization mechanism available in \dunefemdg requires the user to provide some
additional information in the \pyth{Model} class, i.e. 
a scalar jump function of the solution $\phi(\um, \up)$ between two
elements $\elem$ and $\neig$ (\pyth{jump(t,x,U,V)} in the Model) and
a suitable velocity $\vecv$ required to compute the smoothness indicator given in equations
\eqref{shockIndi}:
\begin{python}[Additional methods on Model class used for stabilization (advection problem]
    Model.jump = lambda t,x,U,V: U - V
    Model.velocity = lambda t,x,U: velocity
\end{python}
We now need to construct the operator to include stabilization:
\begin{python}[Setting up the spatial operator to include stabilization]
operator = femDGOperator(Model, space, limiter="default")
U_h      = space.interpolate(Model.U0, name="solution")
operator.applyLimiter(U_h)             # apply limiter to initial solution
\end{python}
A result for \textit{three body problem} including a stabilization is presented
in Fig. \ref{fig:advectionnone} (b) and  Fig. \ref{fig:advectionminmod} (b-d).

Although over- and undershoots are clearly reduced by the stabilization
approach there are still some oscillations clearly visible. They can be
removed by basing the troubled cell indicator on a physicality check, e.g.,
requiring that the solution remains in the interval $[0,1]$. So (possibly instead of
the \pyth{jump,velocity}) attributes we add
\begin{python}[Adding a phyiscality check to the Model (three body advection)]
Model.physical = lambda t,x,U: (
              conditional( U[0]>-1e-8, 1.0, 0.0 ) *
              conditional( U[0]<1.0+1e-8, 1.0, 0.0 ) )
\end{python}
to the code. Another alternative is to use the approach detailed in
\cite{zhang:10,cheng:13} which required to add a tuple with upper and lower
bounds for each component in the solution vector
to the \pyth{Model} and to change the \pyth{limiter} parameter in
the operator constructor:
\begin{python}[Use of spatial operator using reconstruction approach based on scaling of higher moments]
Model.lowerBound = [0] # the scalar quantity should be positive
Model.upperBound = [1] # the scalar quantity should be less then one

operator = femDGOperator(Model, space, limiter="scaling")
\end{python}
The resulting plots using the scaling limiter and simply limiting in cells
where the solution is below zero or above one, are shown in Fig. \ref{fig:advection_phys_scaling}.

\subsubsection{Adaptivity}
We use dynamic local refinement and coarsening of the grid to improve the efficiency of our
simulation. Elements are marked for refinement or coarsening based on an
elementwise constant indicator as discussed in Section
\ref{sec:dgAdaptivity}.
Although this is a more general feature of the \dunefem framework on which
the package is based, adaptivity is such an important tool to improve
efficiency of schemes for the type of problems discussed here, that we will
describe how to use it in some detail:

The grid modification requires an indicator \pyth{indicator} computed between time steps
which provides the information for each cell whether to keep it as it is, refine it, or coarsen it.
To this end \dunefem provides a very simple function
\begin{python}[function for flagging cells for refinement or coarsening]
dune.fem.markNeighbors(indicator, refineTol, coarsenTol, minLevel, maxLevel)
\end{python}
The \pyth{indicator} provides a number $\eta_E\geq 0$ on each element $E$ which is used to
determine if an element is to be refined or coarsened. An element $E$ is
refined, if it's level in the grid hierarchy is less then \pyth{minLevel}
and $\eta_E>$\pyth{refineTol}; it is coarsened if it's level is
greater then \pyth{maxLevel} and $\eta_E<$\pyth{coarsenTol};
otherwise it remains unchanged. In addition if a cell is to be refined all
its neighboring cells are refined as well, so that important structures in
the solution do not move out of refined regions during a time step.
If the initial grid is suitably refined, then one can take
\pyth{maxLevel=0}. The choice for \pyth{minLevel} will depend on the
problem and has to be chosen carefully since increasing \pyth{minLevel} by
one can potentially double the runtime simply due to the reduction in the
time step due to the CFL condition. At the moment we do not provide any
spatially varying time step control. We usually choose
\pyth{coarsenTol} simply as a fraction of \pyth{refineTol}.
In our simulation we use \pyth{coarsenTol=0.2*refineTol}.
This leaves us needing to describe our choice for \pyth{indicator} and
\pyth{refineTol}.

In the results shown here. the indicator $(\eta_E)_E$ is based on the
residual of an auxiliary scalar PDE 
\begin{equation*}
  \partial_t \eta(t,x,U) + \nabla \cdot ( F(t,x,U,\nabla U) )  = S(t,x,U,\nabla U).
\end{equation*}
The function $\eta$, flux $F$ and source $S$ are provided in a subclass \pyth{Indicator}
of the \pyth{Model} class.

\begin{figure}
\centering
  \subfloat[adaptation \& no stabilization]{
    \includegraphics[width=0.48\textwidth]{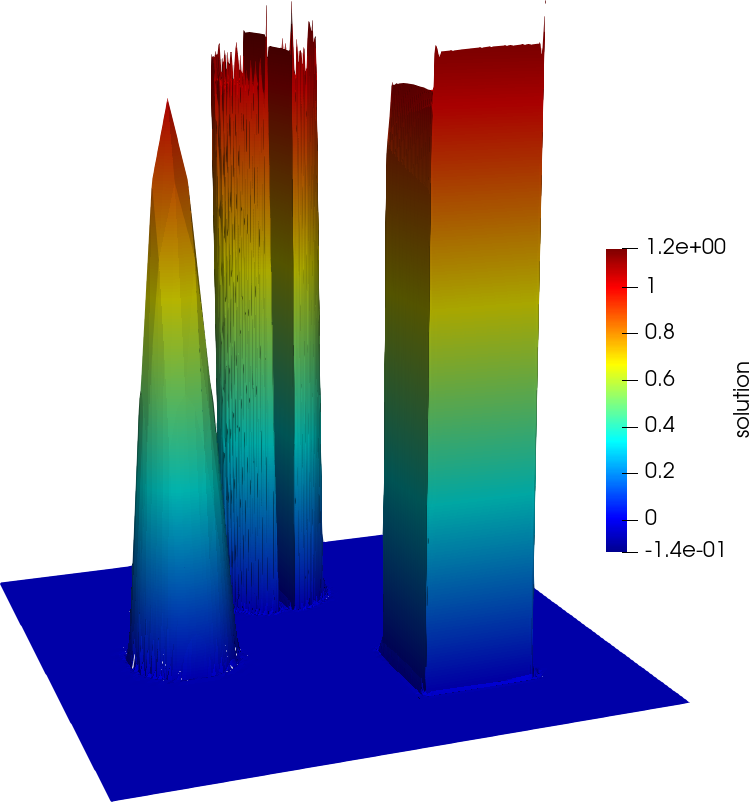}}
  \subfloat[adaptation \& default stabilization]{
    \includegraphics[width=0.48\textwidth]{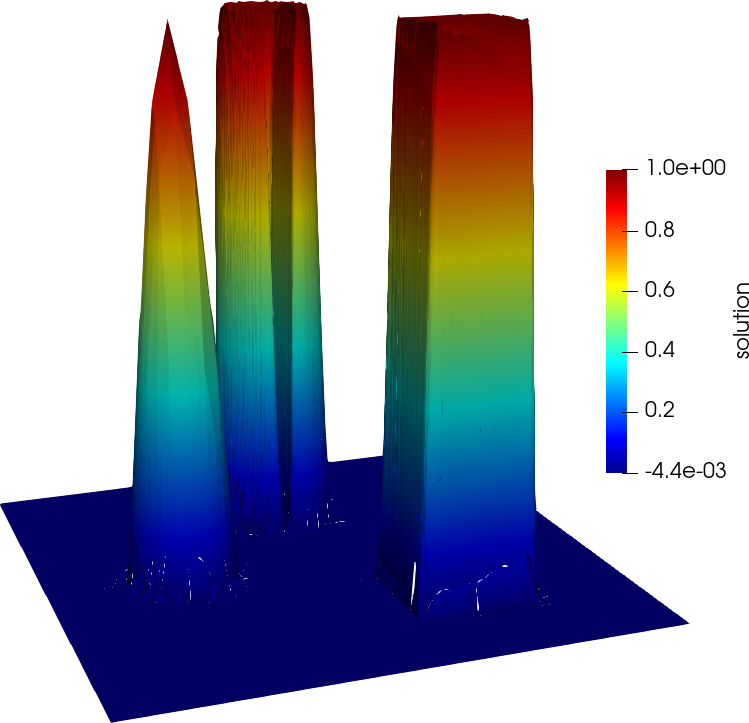}}
\caption{Top view of three body advection problem at time $t=\pi$ using an adaptive
grid. Left without stabilization and right with stabilization.
Color range based on minimum and maximum values of discrete solution.}
\label{fig:advection_adaptive3d}
\end{figure}

\begin{figure}
\centering
  \subfloat[adaptation \& no stabilization]{
    \includegraphics[width=0.48\textwidth]{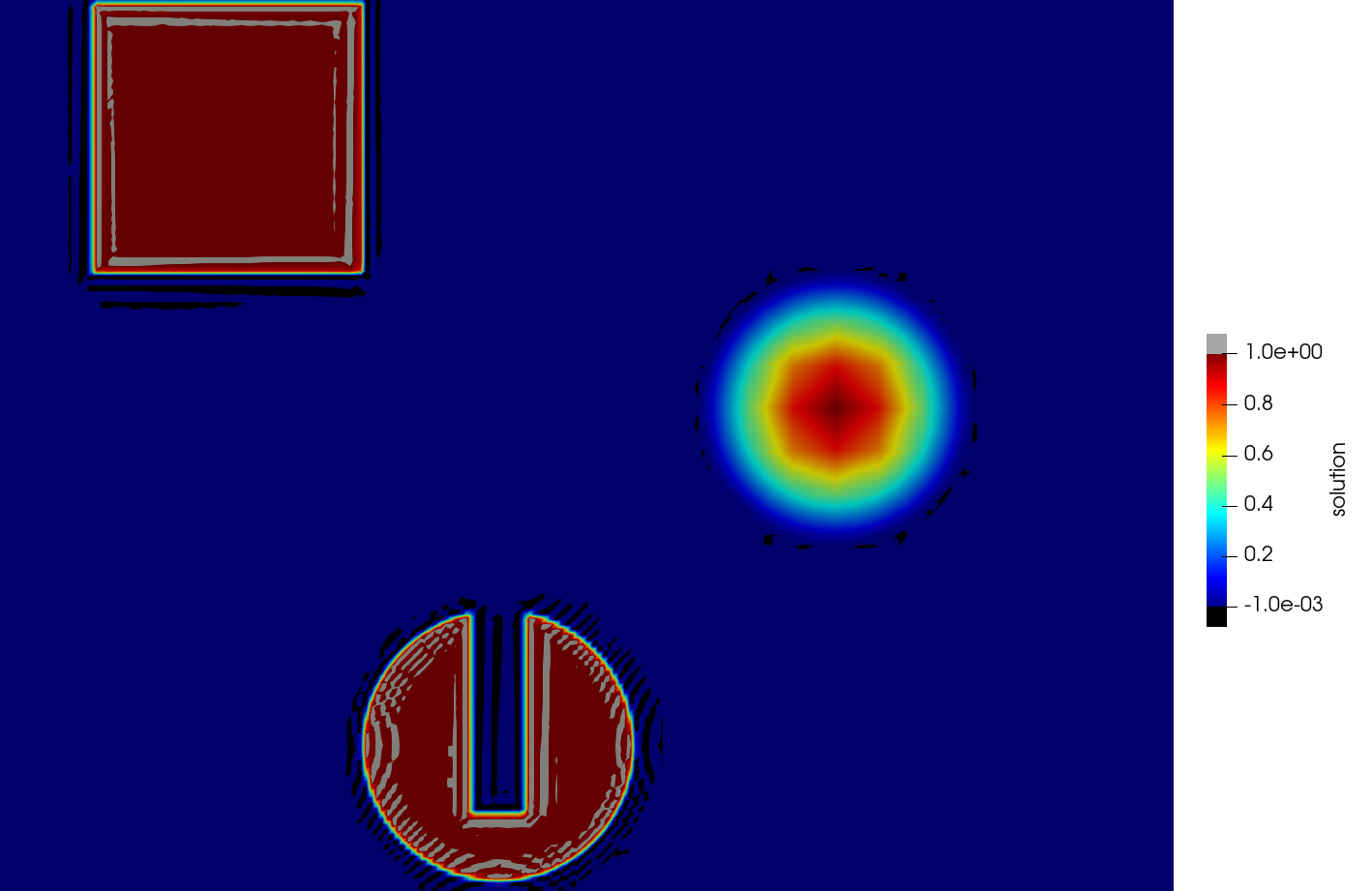}}
  \subfloat[adaptation \& default stabilization]{
    \includegraphics[width=0.48\textwidth]{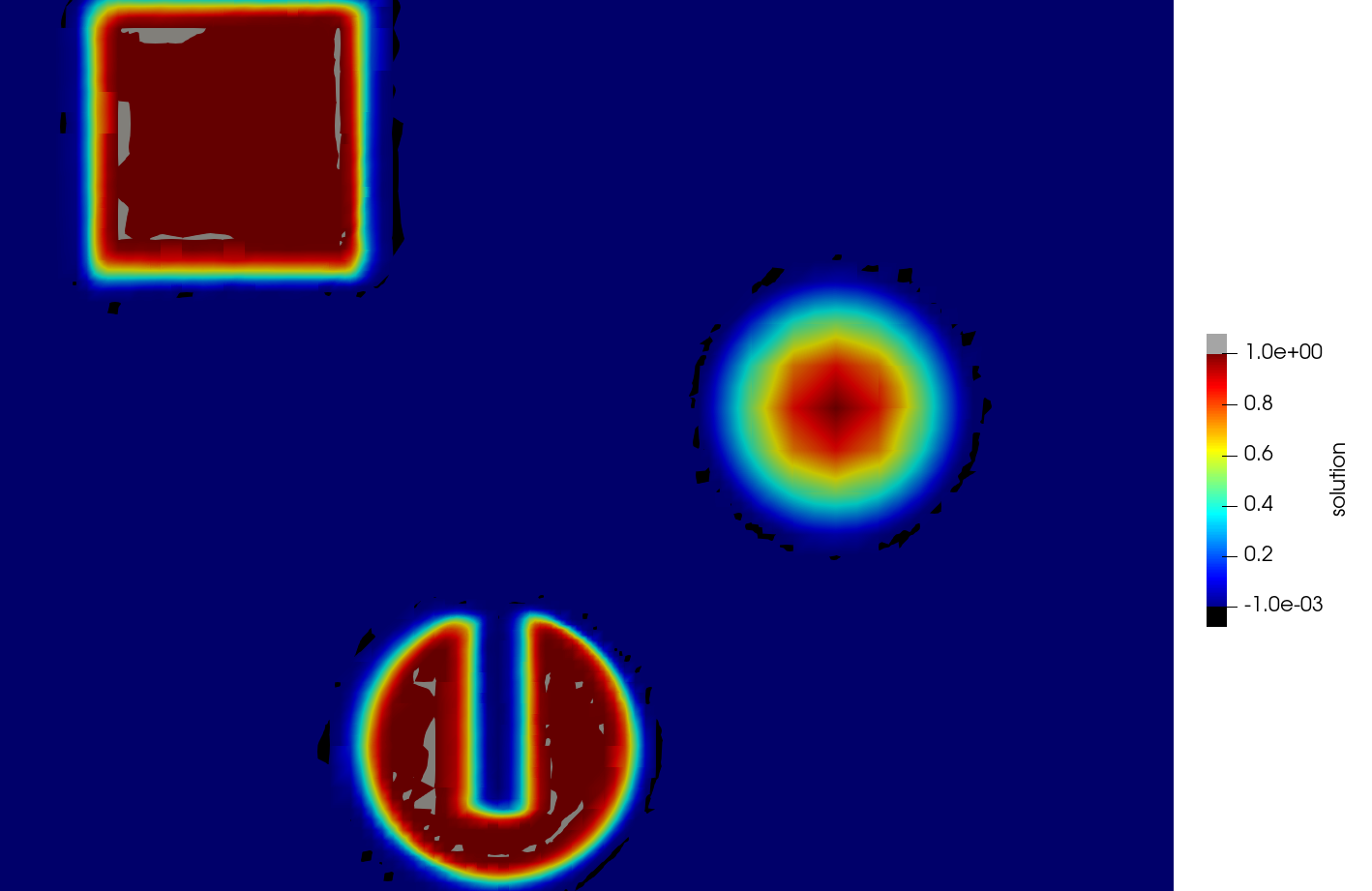}} \\
  \subfloat[adapted grid of a)]{
    \includegraphics[width=0.48\textwidth]{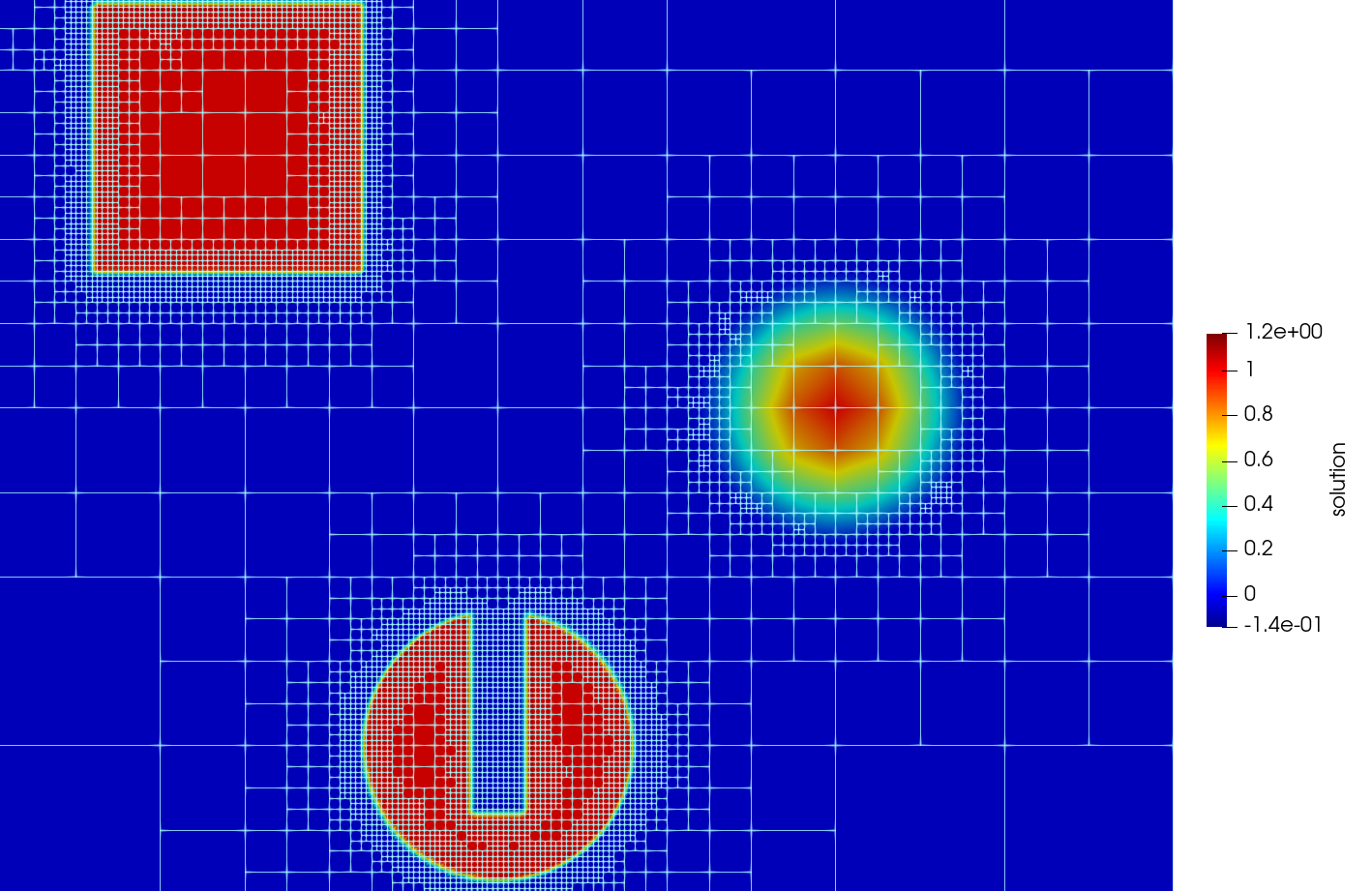}}
  \subfloat[adapted grid of b)]{
    \includegraphics[width=0.48\textwidth]{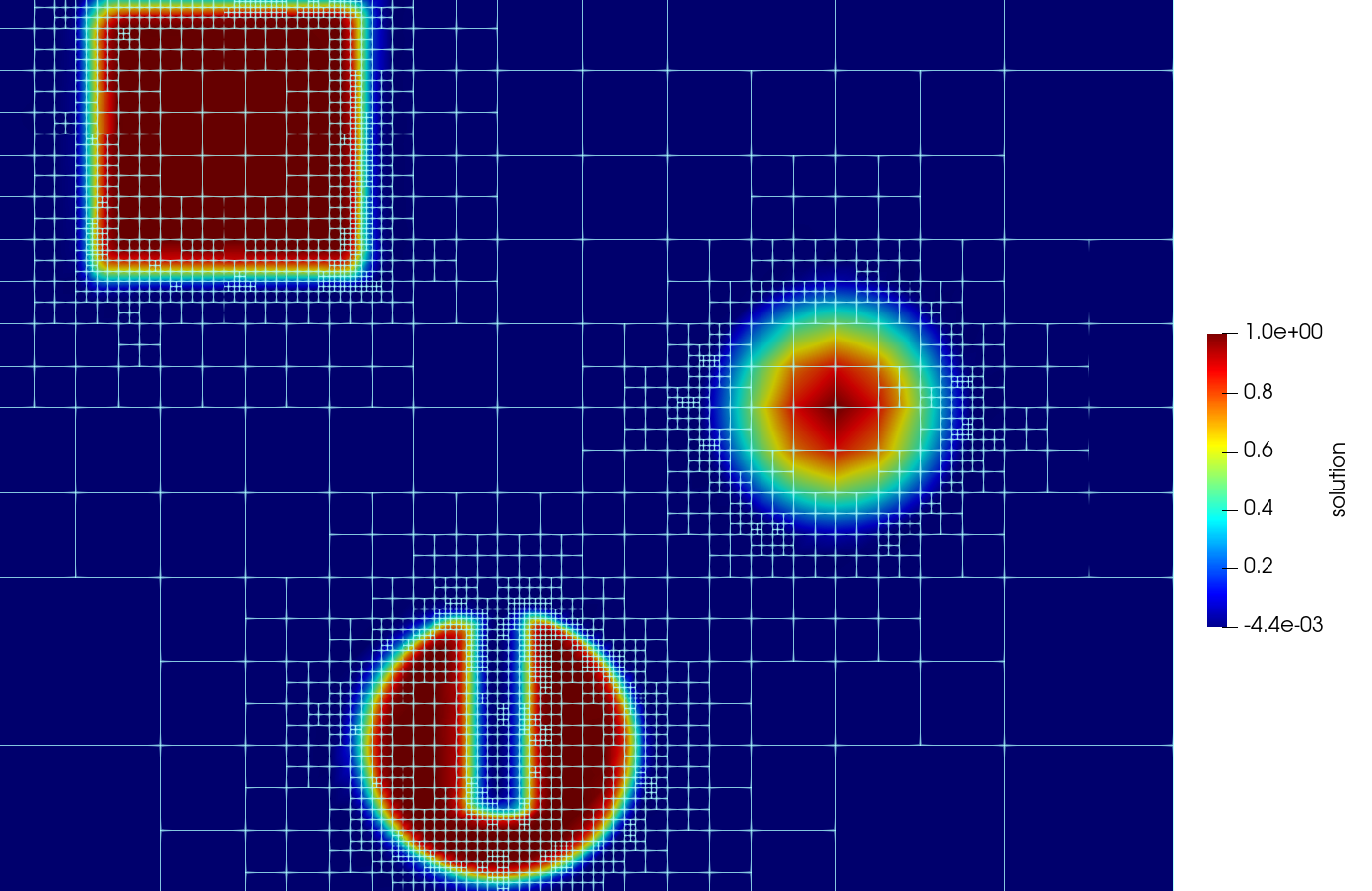}}
\caption{Top view of three body advection problem at time $t=\pi$ using an adaptive
  grid. Left without stabilization and right with stabilization.
Top shows solution with values above $1+10^{-3}$ or below $-10^{-3}$ are colored in grey and black,
respectively. Bottom row shows corresponding grid with solution colored
according to their respective minimum and maximum values.}
\label{fig:advection_adaptivegrid}
\end{figure}

For the advection problem we will simply use the original PDE:
\begin{python}[Adding method to the Model (three body advection) used for residual indicator]
class Indicator:
    def eta(t,x,U):  return U[0]
    def F(t,x,U,DU): return Model.F_c(t,x,U)[0]
    def S(t,x,U,DU): return 0
Model.Indicator = Indicator
\end{python}
The indicator is now computed by applying an operator
mapping the DG space onto a finite volume space resulting an elementwise
constant value for the indicator which we can use to refine/coarsen the
grid:
\begin{python}[Setting up the residual indicator]
# previous version used three global refinement steps
maxLevel = 3    # maximal allowed level for grid refinement
un = U_h.copy() # to store solution at previous time

from dune.fem.space import finiteVolume
from dune.ufl import Constant
from dune.fem import markNeighbors, adapt
from dune.fem.operator import galerkin

indicatorSpace = finiteVolume( gridView )
indicator = indicatorSpace.interpolate(0,name="indicator")
u, phi = TrialFunction(space), TestFunction(indicatorSpace)
dt, t  = Constant(1,"dt"), Constant(0,"t")
x, n   = SpatialCoordinate(space), FacetNormal(space)
hT, he = MaxCellEdgeLength(space), avg( CellVolume(space) ) / FacetArea(space)

eta, F, S = Model.Indicator.eta, Model.Indicator.F, Model.Indicator.S
eta_new, eta_old = eta(t+dt,x,u), eta(t,x,un)
etaMid = ( eta(t,x,un) + eta(t+dt,x,u) ) / 2
FMid   = ( F(t,x,un,grad(un)) + F(t+dt,x,u,grad(u)) ) / 2
SMid   = ( S(t,x,un,grad(un)) + S(t+dt,x,u,grad(u)) ) / 2
Rvol   = (eta_new-eta_old)/dt + div(FMid) - SMid
estimator = hT**2 * Rvol**2 * phi * dx  +\
            he    * inner(jump(FMid), n('+'))**2 * avg(phi) * dS +\
            1/he  * jump(etaMid)**2 * avg(phi) * dS
residualOperator = galerkin(estimator)
\end{python}
In the following we refine the initial grid to accurately resolve the
initial condition. We also use the initial indicator to fix a suitable
value for the \pyth{refineTol}:
\begin{python}[Initial grid adaptation]
maxSize = gridView.size(0)*2**(gridView.dimension*maxLevel)

# initial refinement
for i in range(maxLevel+1):
    un.assign(U_h)
    dt = stepper(U_h)
    residualOperator.model.dt = dt
    residualOperator(U_h,indicator)
    timeTol = gridView.comm.sum( sum(indicator.dofVector) ) / Model.endTime / maxSize
    hTol = timeTol * dt
    markNeighbors(indicator, refineTolerance=hTol, coarsenTolerance=0.2*hTol,
                  minLevel=0, maxLevel=maxLevel)
    adapt(U_h)

    U_h.interpolate(Model.U0)
    operator.applyLimiter(U_h)
\end{python}
Within the time loop we now also need to add the marking and refinement
methods, we repeat the full time loop here:
\begin{python}[Time loop including dynamic grid adaptation]
t = 0
while t < Model.endTime:
    un.assign(U_h)
    dt = stepper(U_h)
    residualOperator.model.dt = dt
    residualOperator(U_h,indicator)
    hTol = timeTol * dt
    markNeighbors(indicator, refineTolerance=hTol, coarsenTolerance=0.1*hTol,
                  minLevel=0, maxLevel=maxLevel)
    adapt(U_h)
    t += dt
\end{python}
Results on a dynamically adapted grid with and without stabilization are
shown in Fig. \ref{fig:advection_adaptive3d} and Fig. \ref{fig:advection_adaptivegrid}.

\subsection{Compressible Euler and Navier-Stokes equations}
\label{sec:euler}
We continue our presentation with a system of evolution equations.
We focus on simulations of the compressible Euler (and later Navier-Stokes)
equations. We show results for two standard test cases: the interaction of a
shock with a low density region and the simulation of a
Kelvin-Helmholtz instability with a density jump.
\subsubsection{Model description}
We start with the methods needed to describe the PDE, the local Lax-Friedrichs flux and
time step control, the troubled cell indicator, and the residual indicator,
All the methods were already discussed in the previous section:
\begin{python}[Full Model class for Euler equations of gas dynamics]
dim=2
class Model:
    gamma = 1.4
    dimRange = dim+2
    # helper function
    def toPrim(U):
        v = as_vector( [U[i]/U[0] for i in range(1,dim+1)] )
        kin = dot(v,v) * U[0] / 2
        pressure = (Model.gamma-1)*(U[dim+1]-kin)
        return U[0], v, pressure
    def toCons(V):
        m = as_vector( [V[i]*V[0] for i in range(1,dim+1)] )
        kin = dot(m,m) / V[0] / 2
        rE = V[dim+1]/(Model.gamma-1) + kin
        return as_vector( [V[0],*m,rE] )

    # interface methods for model
    def F_c(t,x,U):
        rho, v, p = Model.toPrim(U)
        v = numpy.array(v)
        res = numpy.vstack([ rho*v,
                             rho*numpy.outer(v,v) + p*numpy.eye(dim),
                             (U[dim+1]+p)*v ])
        return as_matrix(res)
    # for local Lax-Friedrichs flux
    def maxWaveSpeed(t,x,U,n):
        rho, v, p = Model.toPrim(U)
        return abs(dot(v,n)) + sqrt(Model.gamma*p/rho)

    # for troubled cell indicator: we use the jump of the pressure
    def velocity(t,x,U):
        _, v ,_ = Model.toPrim(U)
        return v
    def jump(t,x,U,V):
        _,_, pU = Model.toPrim(U)
        _,_, pV = Model.toPrim(V)
        return (pU - pV)/(0.5*(pU + pV))
    # negative density/pressure are unphysical
    def physical(t,x,U):
        rho, _, p = Model.toPrim(U)
        return conditional( rho>1e-8, conditional( p>1e-8 , 1, 0 ), 0 )

    # for the residual indicator using the entropy equation
    class Indicator:
        def eta(t,x,U):
            _,_, p = Model.toPrim(U)
            return U[0]*ln(p/U[0]**Model.gamma)
        def F(t,x,U,DU):
            s = Model.Indicator.eta(t,x,U)
            _,v,_ = Model.toPrim(U)
            return v*s
        def S(t,x,U,DU):
            return 0
\end{python}
The next step is to fix the initial conditions and the end time for the two
problems we want to study. First for the shock bubble problem
\begin{python}[Adding initial conditions for shock bubble problem to model class]
    # shock
    gam = 1.4
    pinf, rinf = 5, ( 1-gam + (gam+1)*pinf )/( (gam+1) + (gam-1)*pinf )
    vinf = (1.0/sqrt(gam)) * (pinf - 1.)/ sqrt( 0.5*((gam+1)/gam) * pinf + 0.5*(gam-1)/gam);
    Ul = Model.toCons( [rinf,vinf]+(dim-1)*[0]+[pinf] )
    Ur = Model.toCons( [1]+dim*[0]+[1] )
    # bubble
    center, R2 = 0.5, 0.2**2
    bubble = Model.toCons( [0.1]+dim*[0]+[1] )
    Model.U0 = conditional( x[0]<-0.25, Ul, conditional( dot(x,x)<R2, bubble, Ur) )
    Model.endTime = 0.5
\end{python}
To complete the description of the problem we need to define boundary
conditions. For the advection problem we used Dirichlet boundary conditions
which are used as second state for the numerical flux over the boundary
segments. For this problem we will use Dirichlet boundary conditions on the
left and right boundary but want to use no flow boundary conditions on
the top and bottom boundary (which will have an identifier of $3$ on this
domain):
\begin{python}[Adding boundary conditions for shock bubble problem to Model class]
    def noFlowFlux(u,n):
        _, _, p = Model.toPrim(u)
        return as_vector([0]+[p*c for c in n]+[0])
    Model.boundary = {1: lambda t,x,u: Ul,
                      2: lambda t,x,u: Ur,
                      3: lambda t,x,u,n: noFlowFlux(u,n)}
    Model.domain = (reader.dgf, "shockbubble"+str(dim)+"d.dgf")
\end{python}
Next we describe initial conditions and boundary conditions for the
Kelvin-Helmholtz instability between two layers with a density jump
where we use periodic boundary conditions in
the horizontal direction and reflective boundary conditions on the vertical
boundaries:
\begin{python}[Initial and boundary conditions for Kelvin-Helmholtz problem]
    sigma = 0.05/sqrt(2)
    rho,pres = conditional( abs(x[1]-0.5)<0.25,2,1), 2.5
    u        = conditional( abs(x[1]-0.5)<0.25,0.5,-0.5)
    v        = 0.1*sin(4*pi*x[0])*( exp(-(x[1]-0.25)**2/(2*sigma**2)) )
    Model.U0 = Model.toCons([rho,u,v,pres])
    def reflect(U,n):
        n,m = as_vector(n), as_vector([U[1],U[2]])
        mref = m - 2*dot(m,n)*n
        return as_vector([U[0],*mref,U[3]])
    Model.boundary = {3: lambda t,x,U: reflect(U,[0,-1]),
                      4: lambda t,x,U: reflect(U,[0,1])}
    Model.domain = (reader.dgf, "kh.dgf")
    Model.endTime = 1.5
\end{python}
Now that we have set up the model class, the code presented for the
advection problem for evolving the system and adapting the grid can remain
unchanged. 

\begin{figure}
\centering
\includegraphics[width=0.48\textwidth]{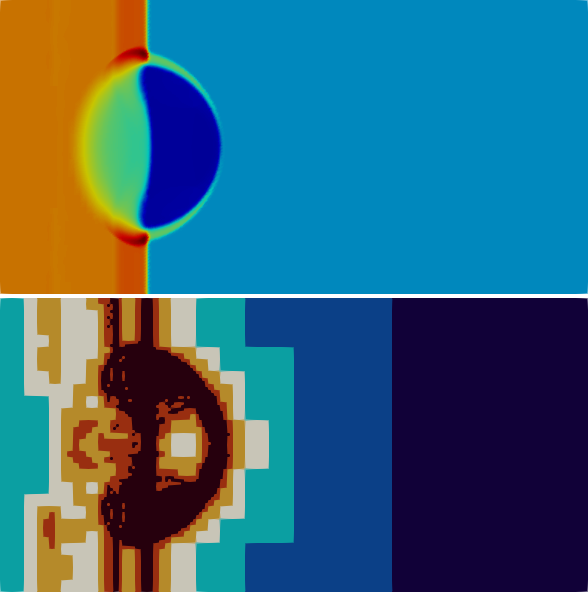}
\includegraphics[width=0.48\textwidth]{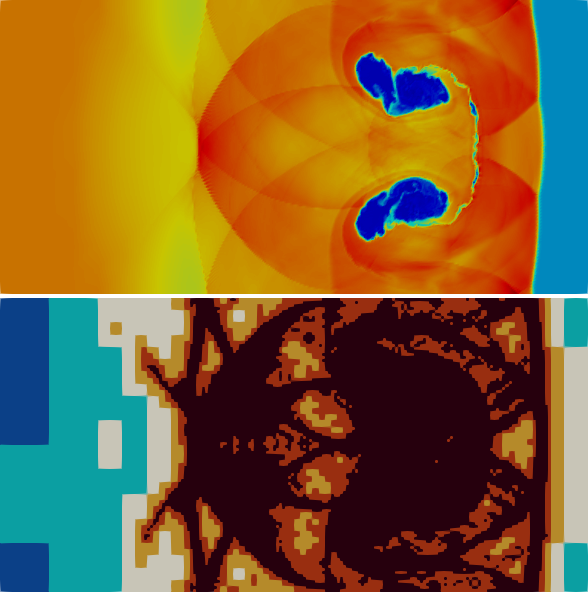}
\caption{Shock bubble (actually column) interaction problem at $t=0.1$ (left) and $t=0.5$
(right). Top figure shows the
density and bottom figure the levels of the dynamically adapted grid.}
\label{fig:sb2d}
\end{figure}

\begin{figure}
\centering
\includegraphics[width=0.98\textwidth]{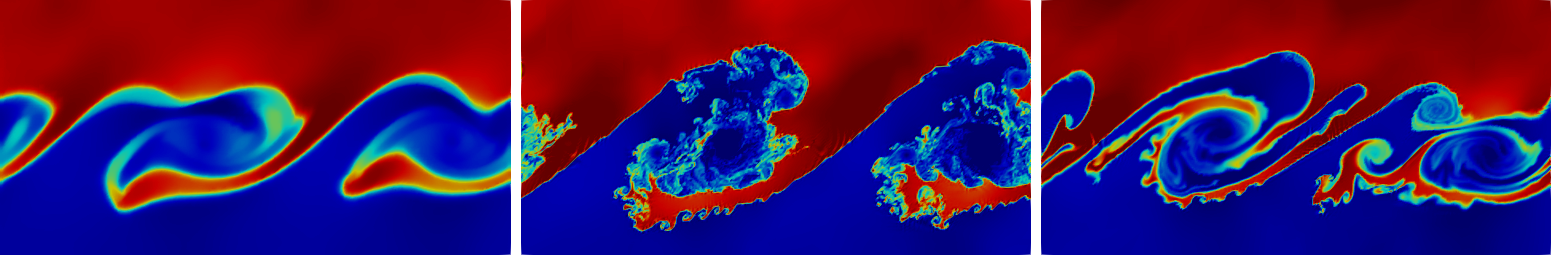}
\includegraphics[width=0.98\textwidth]{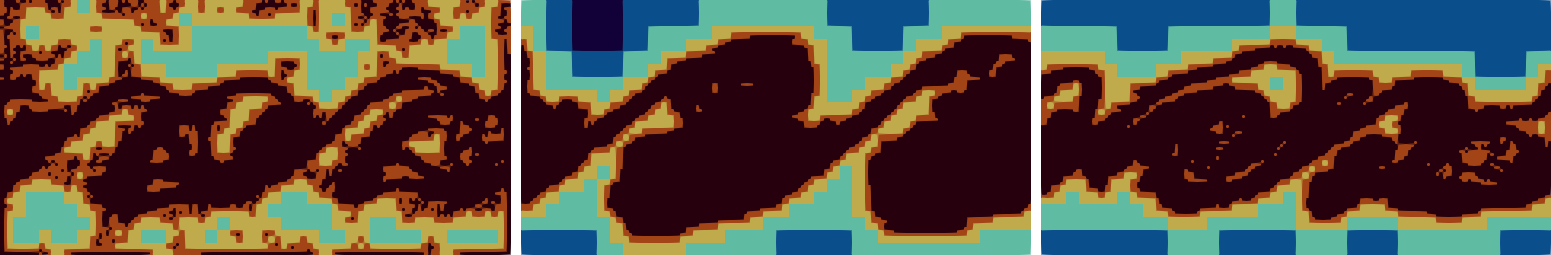}
\caption{Density (top) and grid levels (bottom)
for Kelvin-Helmholtz instability at time $t=1.5$. In the initial setup the
density in the upper layer is $2$ and $1$ in the lower layer.
The middle figure shows the default setting for the troubled cell
indicator which uses a tolerance of $1$. The right figure shows results
using a reduced tolerance of $0.2$ so that more cells are marked.
On the left we show results obtained using the indicator from \cite{Kloeckner} based
on the modal expansion of the density. Details on how this indicator is
added to the code is discussed in Section~\ref{sec:userdeftroubledcell}.
}
\label{fig:khdensity_level}
\end{figure}

Results for both test cases on locally adapted grids are shown in
Fig.~\ref{fig:sb2d} and the middle column of Fig. \ref{fig:khdensity_level}.
The default setting works very well for the shock bubble interaction problem.
It turns out that in the Kelvin-Helmholtz case the stabilization is almost
completely determined by the physicality check since the smoothness indicator
shown here is based on the pressure which in this case is continuous over the
discontinuity. To increase the stabilization of the method it is either
possible to reduce the tolerance in the troubled cell indicator
or to use a different
smoothness indicator all together, which is discussed in
Section~\ref{sec:userdeftroubledcell}.
Using the default setting for the indicator, very fine structures appear
with considerable under- and overshoots developing as shown in the middle figure of
Fig. \ref{fig:khdensity_level}. The minimum and
maximum density are around $0.6$ and $2.6$, respectively.
After increasing the sensitivity of the smoothness indicator, by passing a suitable parameter to the
constructor of the operator to reduce the tolerance (using $TOL=0.2$ here)
the under- and overshoots are reduced to $0.8$ and $2.4$ and some of the
fine structure has been removed as shown on the right of
Fig. \ref{fig:khdensity_level}.
On the left of the same figure we show results form a simulation using the indicator from
\cite{Kloeckner} based on the modal expansion of the density. Here minimum and maximum densities
are $0.95$ and $2.1$, respectively.
More details of this indicator are provided in Section~\ref{sec:userdeftroubledcell}.
As reference for the Navier-Stokes
simulation shown on the right of Fig.~\ref{fig:khns} density throughout
the simulation was in the range $0.97$ and $2.1$. This is discussed further
in Section~\ref{sec:userdeftroubledcell}


\subsubsection{Adding source terms}
\begin{figure}
\centering
\includegraphics[width=0.48\textwidth]{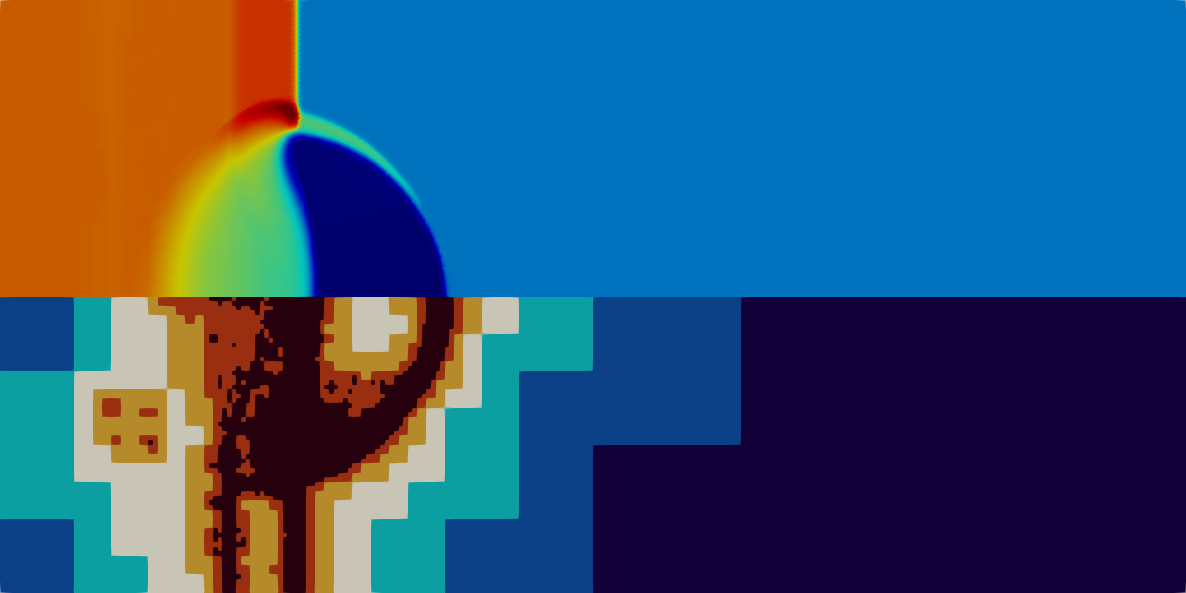}
\includegraphics[width=0.48\textwidth]{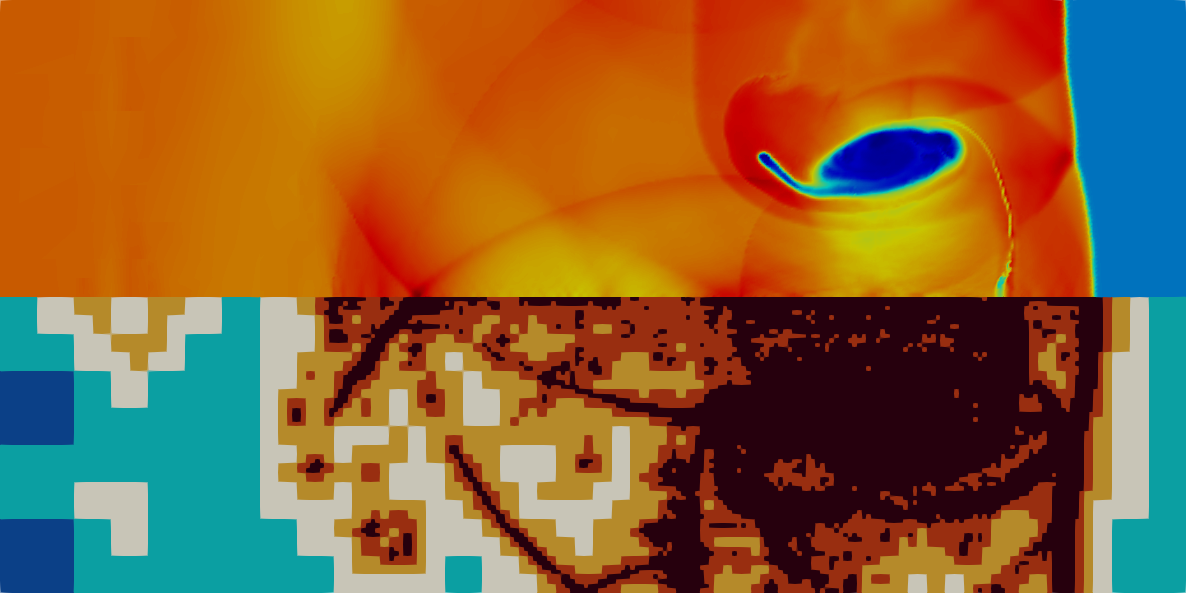}
\caption{Shock bubble interaction problem in cylindrical coordinate at time
$t=0.1$ and $t=0.5$. Density is shown in the top row and the adaptive grid
in the bottom row.}
\label{fig:sbR}
\end{figure}

So far we have in fact simulated the interaction of a shock wave with a
column of low density gas and not in fact a bubble. To do the later would
require to either extend the problem to 3D (discussed in the next section)
or to simulate the problem using cylindrical coordinates. This requires
adding a geometric source term to the right hand side of the Euler
equations. As pointed out in the introduction 
our model can take two types of source term depending upon the desired
treatment in the time stepping scheme. Here we want to treat the source
explicitly so need to add an \pyth{S_e} method to the \pyth{Model} class:
\begin{python}[Adding source term to Model class for Euler equations]
# geometric source term for cylindrical coordinates
def source(t,x,U,DU):
    _, v, p = Model.toPrim(U)
    return as_vector([ - U[0]   *v[1]/x[1],
                       - U[1]   *v[1]/x[1], - U[2]   *v[1]/x[1],
                       -(U[3]+p)*v[1]/x[1] ])
Model.S_e = source
# additional source term for residual indicator
def indicatorSource(t,x,U,DU):
    s = Model.Indicator.eta(t,x,U)
    return - s * U[1]/U[0]/x[1]
Model.Indicator.S = indicatorSource
\end{python}
Results are shown in Fig. \ref{fig:sbR} which show a clear difference in
the structure of the bubble at later time compared to Fig. \ref{fig:sb2d}
which is matched by the structure of the full 3d simulation shown in
Fig. \ref{fig:sb3d}.

\subsubsection{Adding diffusion}
\begin{figure}
\centering
\includegraphics[width=0.98\textwidth]{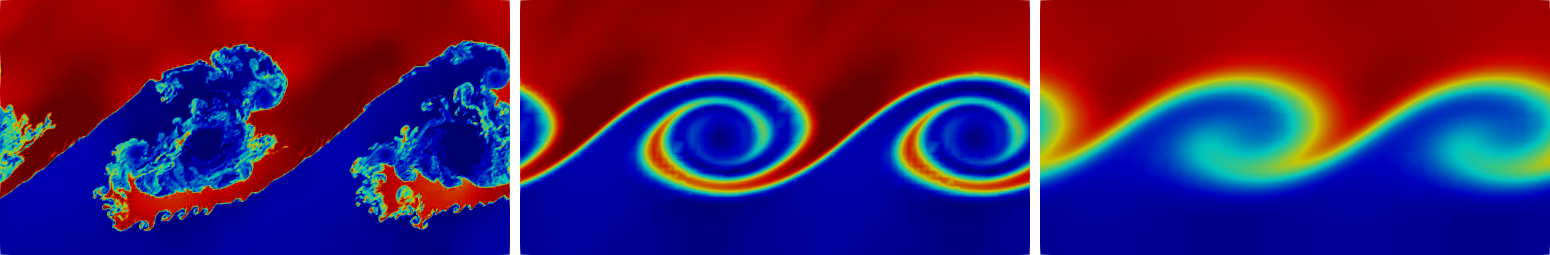}
\includegraphics[width=0.98\textwidth]{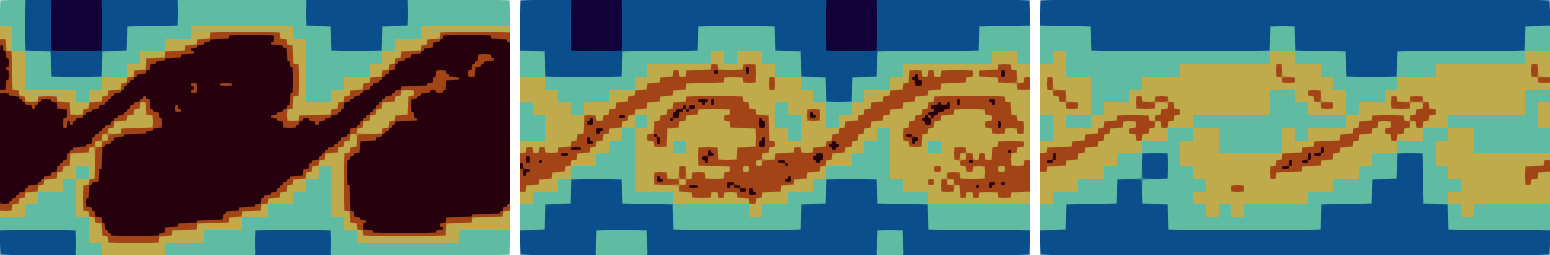}
\caption{Diffusive Kelvin-Helmholtz instability with $\mu=0$ (Euler), $\mu=0.0001$,
and $\mu=0.001$ (from left to right). Top row shows density and bottom row
the grid levels of the adaptive simulation. The smoothing
effect of the viscosity is clearly visible compared to the results shown
for the Euler equations. Consequently, the small scale instabilities are completely
suppressed and a coarser grid is used.}
\label{fig:khns}
\end{figure}

Finally, we discuss the steps needed to add a diffusion term. This requires
adding an additional method to the \pyth{Model} class. So to solve the
compressible Navier-Stokes equations instead of the Euler equations the
following method needs to be added (given a viscosity parameter $\mu$):

\begin{python}[Implementing Navier-Stokes by adding diffusion to Model class for Euler equations]
def F_v(t,x, U, DU):
    assert dim == 2
    Pr = 0.72

    rho, rhou, rhoE = U[0], as_vector([U[j] for j in range(1,dim+1)]), U[dim+1]
    grad_rho = DU[0, :]
    grad_rhou = as_matrix([[DU[j,:] for j in range(1, dim+1)]])[0]
    grad_rhoE = DU[dim+1,:]

    grad_u = as_matrix([[(grad_rhou[j,:]*rho - rhou[j]*grad_rho)/rho**2 for j in range(dim)]])[0]
    grad_E = (grad_rhoE*rho - rhoE*grad_rho)/rho**2

    tau = mu*(grad_u + grad_u.T - 2.0/3.0*tr(grad_u)*Identity(dim))
    K_grad_T = mu*Model.gamma/Pr*(grad_E - dot(rhou, grad_u)/rho)
    return as_matrix([
        [0.0,                                0.0],
        [tau[0,0],                           tau[0,1]],
        [tau[1,0],                           tau[1,1]],
        [dot(tau[0,:], rhou)/rho + K_grad_T[0], dot(tau[1,:], rhou)/rho + K_grad_T[1]] ])
\end{python}
As an example we repeat the simulation of the Kelvin-Helmholtz instability
(see Fig. \ref{fig:khns}). Note that with the default setting for the
stepper, an IMEX scheme is used where the diffusion is treated implicitly
and the advection explicitly with a time step given by the CFL condition.
Consequently, we do not need to make any change to the construction of the
spatial operator and time stepper shown in the code listing on
Page~\pageref{opcode}.

\subsection{User defined smoothness indicator}
\label{sec:userdeftroubledcell}

To exchange the smoothness indicator the construction of the operator
has to be slightly changed. Assuming that the indicator is defined in a
source file \file{modalindicator.hh} which defines a C++ class 
\cpp{ModalIndicator} which takes the C++ type of the
discrete function \pyth{U_h} as template argument. This class needs to be
derived from a pure virtual base class and override a single method.
Then the construction of the operator needs to be changes to
\begin{python}[Changing the smoothness indicator for the stabilization in the spatial operator]
from dune.typeregistry import generateTypeName
from dune.femdg import smoothnessIndicator

# compile and load the module for the smoothness indicator - need the correct C++ type
clsName,includes = generateTypeName("ModalIndicator", U_h)
# the ModelIndicator class has a default constructor (ctor) without arguments
indicator = smoothnessIndicator(clsName, ["modalindicator.hh"]+includes, U_h, ctorArgs=[])
# construct the operator
operator = femDGOperator(Model, space, limiter=["default",indicator])
\end{python}
As an example we use here a smoothness indicator based on studying the
decay properties of the modal expansion of the solution on each cell
following the ideas presented in \cite{Kloeckner}.
For the following implementation we assume that we are using a modal basis
function set orthonormalized over the reference element. Then the C++ code
required to compute the smoothness indicator based on the modal
expansion of the density is given in the next snippets. 
\begin{c++}[C++ class used for user defined smoothness indicator]
template <class DiscreteFunction>
struct ModalIndicator
: public Dune::Fem::TroubledCellIndicatorBase<DiscreteFunction> {
  using LocalFunctionType = DiscreteFunction::LocalFunctionType;
  ModalIndicator () {}
  double operator()( const DiscreteFunction& U,
                     const LocalFunctionType& uEn) const override {
    double modalInd = smoothnessIndicator( uEn );
    return std::abs( modalInd ) > 1e-14?  1. / modalInd : 0.;
  }
\end{c++}
The actual computation of the indicator is carried out in the method
\cpp{smoothnessIndicator} but is slightly too long to include here
directly but is shown in Appendix~\ref{app:modalind}.
It is important to note that it requires very little knowledge of
the \dune programming environment or even C++ since it relies mainly on the
local degrees of freedom vector provided by the argument \cpp{uEn}.
Simulation results for the Kelvin-Helmholtz instability using this
indicator are included in Fig.~\ref{fig:khdensity_level}.


\subsection{User defined time stepping schemes}
\label{sec:userstepper}
The \dunefem package provides a number of standard strong stability preserving
Runge-Kutta (SSP RK) solvers including
explicit, diagonally implicit, and IMEX schemes of degree one to four.
In the literature there is a wide range of additional suitable RK methods
(having low storage or better CFL constants using additional stages for
example). Furthermore, multistep methods can be used. Also there are a
number of other packages providing implementations of timestepping methods.
Since the computationally critical part of a DG method of the type
described here lies in the computation of the spatial operator, the
additional work needed for the timestepper can be carried out on the Python
side with little impact. Furthermore, as pointed out in the introduction, 
it is often desirable to use Python for rapid prototyping and
to then reimplement the finished algorithm in C++ after a first testing
phase to avoid even the slightest impact on performance. The following code
snippet shows how a multistage third order RK method taken from \cite{Ketcheson:08}
can be easily implemented in Python and used to replace the \pyth{stepper}
used so far:
\begin{python}[Implementation of time stepping method on Python side]
class ExplSSP3:
    def __init__(self,stages,op,cfl=0.45):
        self.op     = op
        self.n      = int(sqrt(stages))
        self.stages = self.n*self.n
        self.r      = self.stages-self.n
        self.q2     = op.space.interpolate(op.space.dimRange*[0],name="q2")
        self.tmp    = self.q2.copy()
        self.cfl    = cfl * stages*(1-1/self.n)
        self.dt     = None
    def c(self,i):
        return (i-1)/(self.n*self.n-self.n) \
               if i<=(self.n+2)*(self.n-1)/2+1 \
               else (i-self.n-1)/(self.n*self.n-self.n)
    def __call__(self,u,dt=None):
        if dt is None and self.dt is None:
            self.op.stepTime(0,0)
            self.op(u, self.tmp)
            dt = self.op.timeStepEstimate[0]*self.cfl
        elif dt is None:
            dt = self.dt
        self.dt = 1e10
        fac = dt/self.r
        i = 1
        while i <= (self.n-1)*(self.n-2)/2:
            self.op.stepTime(self.c(i),dt)
            self.op(u,self.tmp)
            self.dt = min(self.dt, self.op.timeStepEstimate[0]*self.cfl)
            u.axpy(fac, self.tmp)
            i += 1
        self.q2.assign(u)
        while i <= self.n*(self.n+1)/2:
            self.op.stepTime(self.c(i),dt)
            self.op(u,self.tmp)
            self.dt = min(self.dt, self.op.timeStepEstimate[0]*self.cfl)
            u.axpy(fac, self.tmp)
            i += 1
        u.as_numpy[:] *= (self.n-1)/(2*self.n-1)
        u.axpy(self.n/(2*self.n-1), self.q2)
        while i <= self.stages:
            self.op.stepTime(self.c(i),dt)
            self.op(u,self.tmp)
            self.dt = min(self.dt, self.op.timeStepEstimate[0]*self.cfl)
            u.axpy(fac, self.tmp)
            i += 1
        self.op.applyLimiter( u )
        self.op.stepTime(0,0)
        return dt

# use a four stage version of this stepper
stepper = ExplSSP3(4,operator)
\end{python}
Again the other parts of the code can remain unchanged.
Results with this time stepping scheme are included in some of the
comparisons shown in the next section
(see Fig. \ref{fig:2vwsmooth} and Fig. \ref{fig:sod}).

\subsection{Different grids and spaces}
\label{sec:gridspace}
\begin{figure}
\centering
\includegraphics[width=0.485\textwidth]{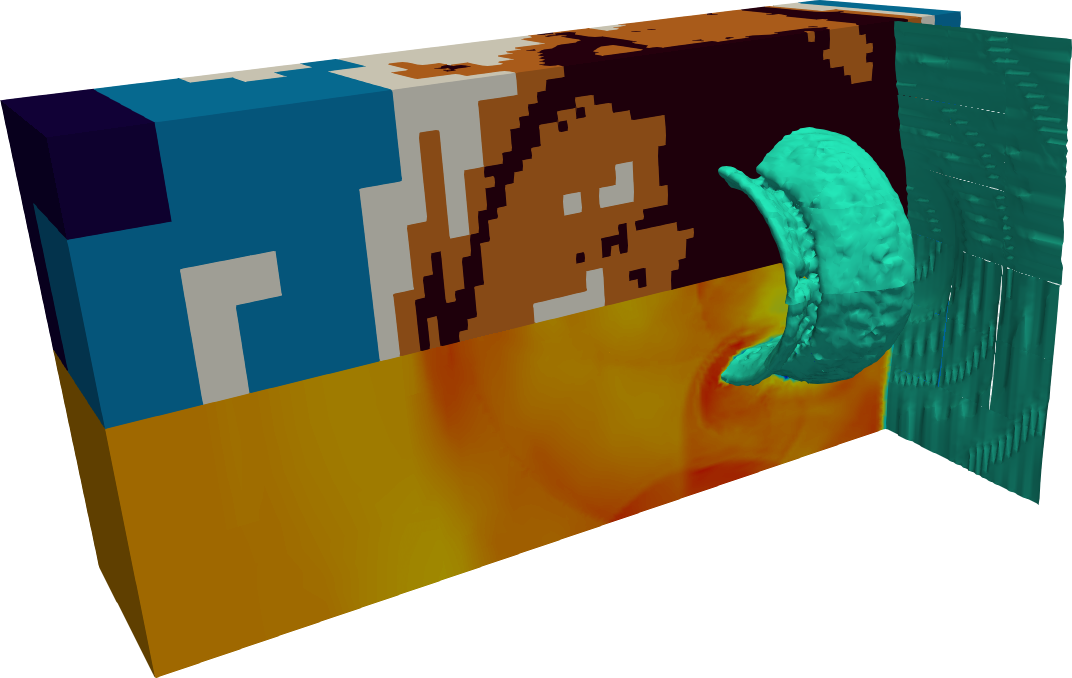}
\includegraphics[width=0.485\textwidth]{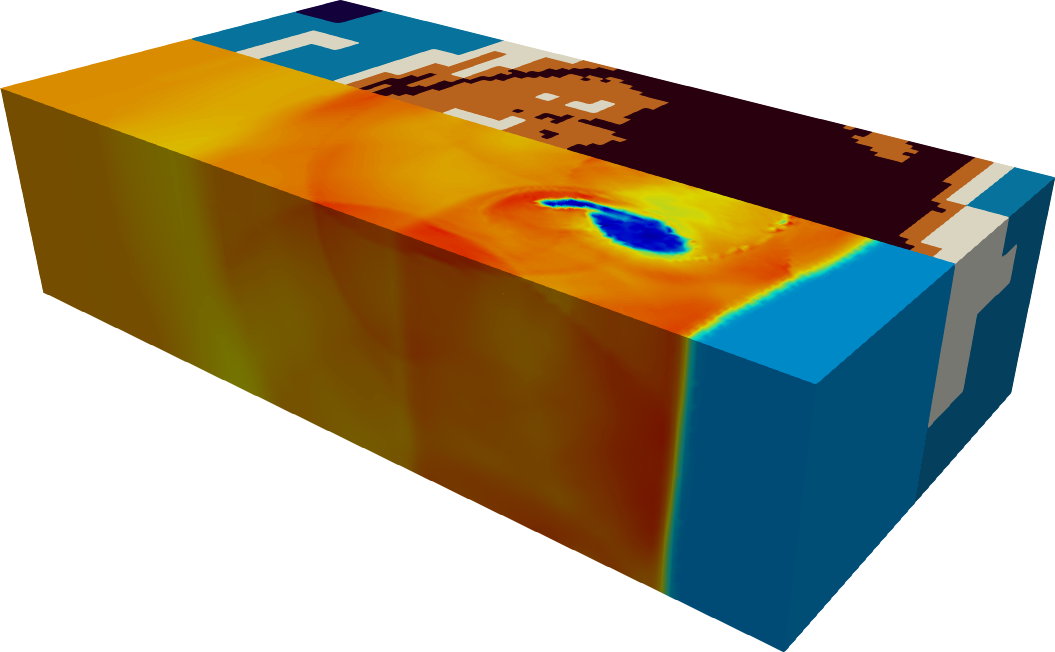}
\caption{Adaptive 3D simulation of shock bubble interaction problem shown
from two different sides. The grid levels and the density distribution at
time $t=0.5$ is shown together with the isosurface $\rho=1.5$.
The simulation was performed on a workstation using 8 processors. The
corresponding globally refined grid would have contained 1.5M elements
resulting in 52M degrees of freedom. The grid shown here consists of
150.000 elements with about 5.2M degrees of freedom.}
\label{fig:sb3d}
\end{figure}

One of the strengths of the \dune framework on which we are basing the
software presented here, is that it can handle many different types of grid
structures.
Performing a 3D simulation can be as simple as changing the \pyth{domain}
attribute in the \pyth{Model} class (Fig. \ref{fig:sb3d}).

\begin{figure}
\centering
\includegraphics[width=0.7\textwidth]{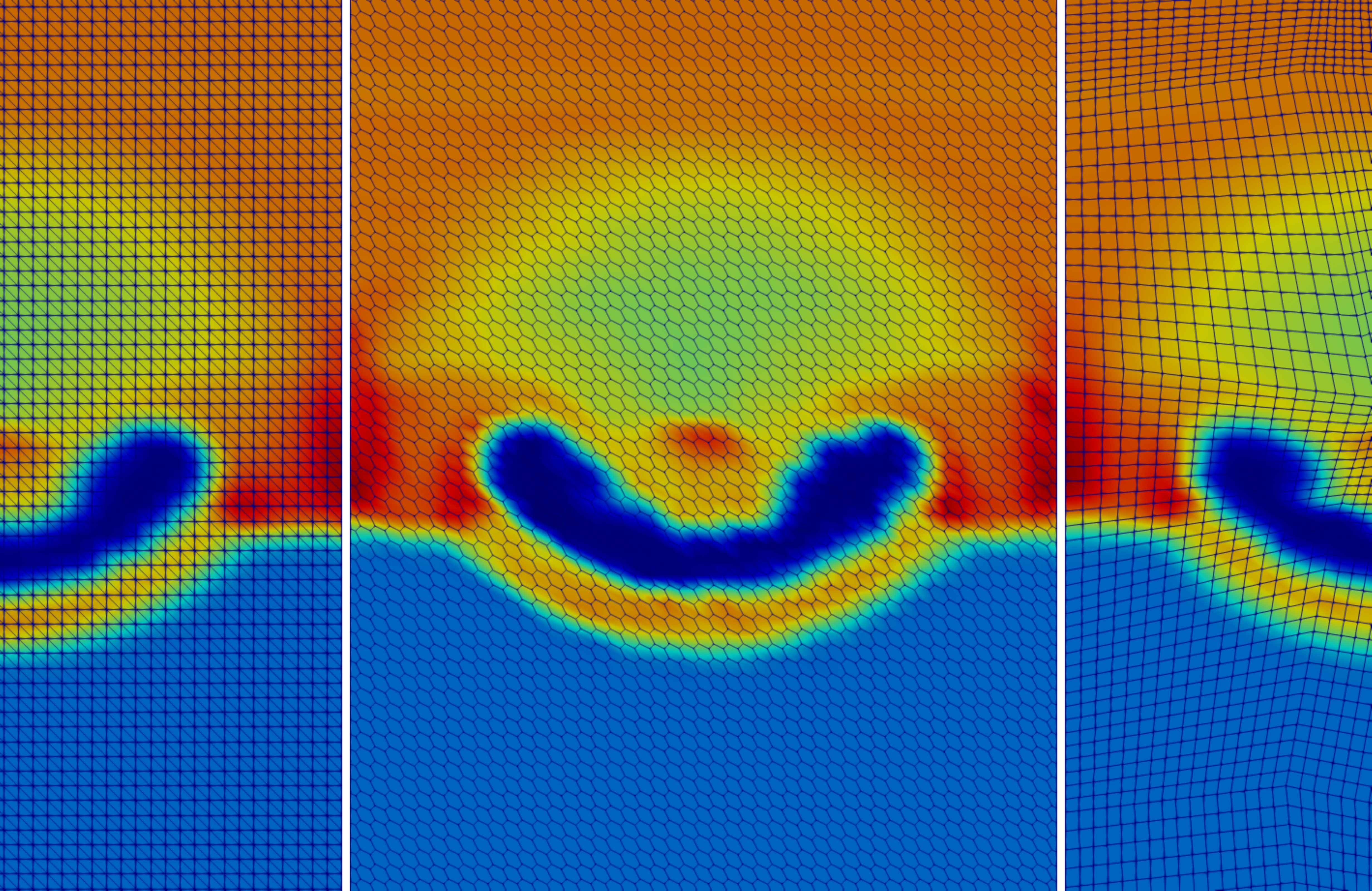}
\caption{Shock bubble interaction problem at $t=0.1$ using different grid
structures. A structured simplex grid (left), a non affine cube grid
(right), a polygonal grid (middle) consisting of the dual of the simplex
grid on the left.}
\label{fig:sbgrids}
\end{figure}

It is also straightforward to perform the simulations on for example a simplicial grid
instead of the cube grid used so far:
\begin{python}[Changing the grid structure (simplex grid)]
from dune.alugrid import aluSimplexGrid as grid

gridView = view( grid( Model.domain ) )
gridView.hierarchicalGrid.globalRefine(3)   # refine a possibly coarse initial grid
\end{python}
It is even possible to use a grid consisting of general polygonal elements,
by simply importing the correct grid implementation:
\enlargethispage{0.2cm}
\begin{python}[Changing the grid structure (polyhedral grid)]
from dune.polygongrid import polygonGrid as grid
\end{python}
A wide range of other grid types are available and a recent overview is given in 
\cite{dunereview:20} some examples are shown in Fig. \ref{fig:sbgrids}.

As pointed out in the previous section, the choice of the basis function
set used to represent the discrete solution can strongly influence the
efficiency of the simulation. So far we have used an orthonormal basis
function set for the polynomial space over the reference cube $[0,1]^d$.
If the grid elements are all affine mapping of $[0,1]^d$ (i.e.
parallelograms) then this is good choice, since it has the minimal number
of degrees of freedom for a desired approximation accuracy while the mass
matrix will be a very simple diagonal matrix on all elements.
These properties always hold for simplicial elements when an orthonormal
polynomial basis over the reference simplex is used. As soon as the mapping
between the reference element and a given element in the grid becomes
non affine, both properties can be lost. To achieve the right approximation properties
in this case, it might be necessary, to use a tensor
product polynomial space, increasing the number of degrees of freedom per
element considerably. Also an orthonormal set of basis function over
the reference cube, will still lead to a dense mass matrix. This lead to a
significant reduction in the efficiency of the method, if the local mass matrix on each
element can not be stored due to memory restrictions.
A possible solution to this problem is to use a Lagrange type set of basis functions with
interpolation points coinciding with a quadrature rule over the reference
cube. We discussed this approach in some detail in
Section~\ref{seq:discretization_spatial}. There we mentioned two choices
for such quadrautre: the use of tensor product Gauss-Legendre rules which
are optimal with respect to accuracy. 
In the software framework presented here, it is straightforward to switch to
this representation of the discrete space by replacing the construction of
the \pyth{space} object by
\begin{python}[Changing the discrete space (dg space using Lagrange type basis functions)]
from dune.fem.space import dglagrange
space = dglagrange( gridView, dimRange=Model.dimRange, order=4, pointType="gauss" )
\end{python}
If the \pyth{operator} is constructed using this space, the suitable
Gaussian quadrature is chosen automatically. Note that this space is only
well defined over a grid consisting of cubes.
A second common choice, which corresponds to an underintegration of the mass
matrix but faster evaluation of surface integrals, is to use 
Lobatto-Gauss-Legendre (LGL) quadrature rules.
Again switching to this Lagrange point set is straightforward
\begin{python}[Changing the discrete space (dg space using spectral basis functions)]
space = dglagrange(gridView, dimRange=Model.dimRange, order=4, pointType="lobatto")
\end{python}
and again using this space in the construction of the spatial operator will
result in the correct LGL quadrature being used.

\begin{figure}
\includegraphics[width=0.475\textwidth]{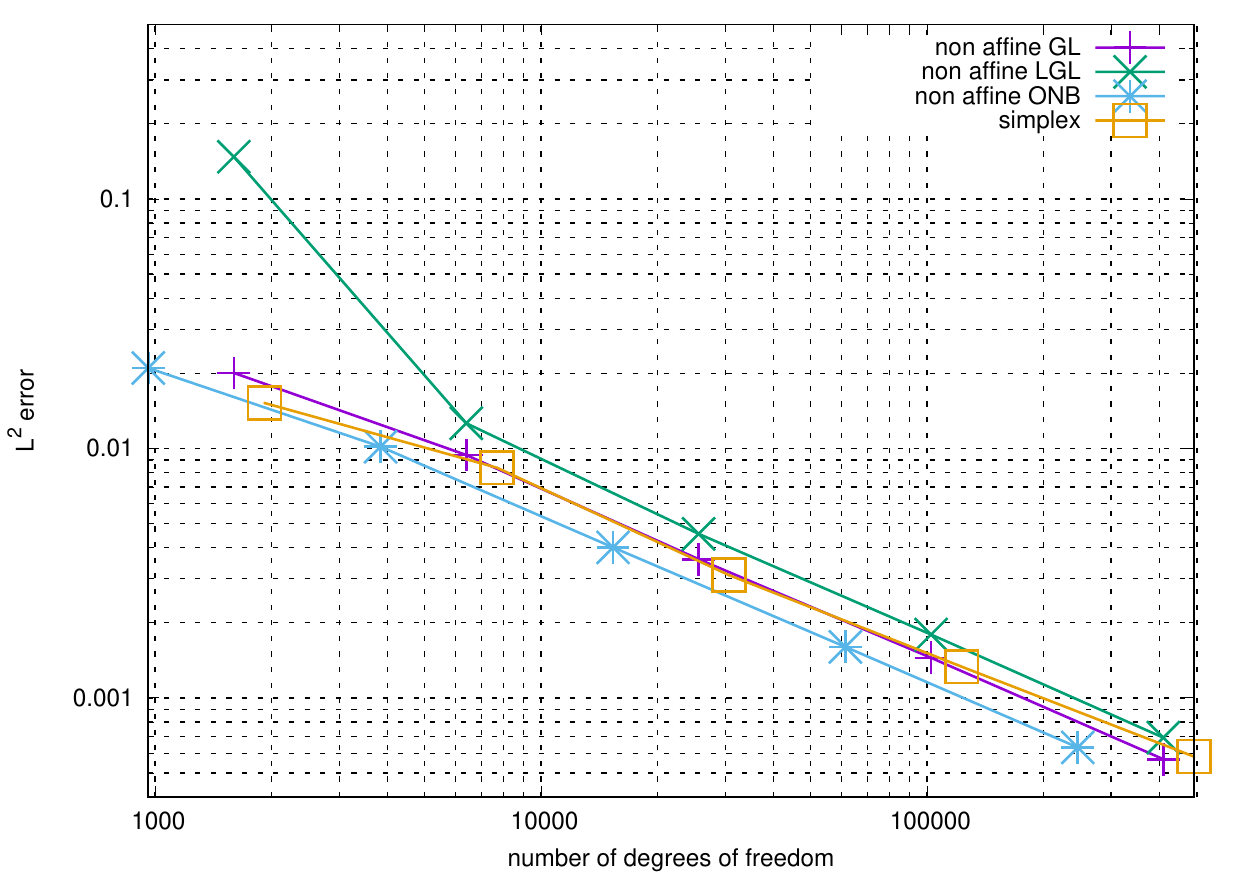}
\includegraphics[width=0.475\textwidth]{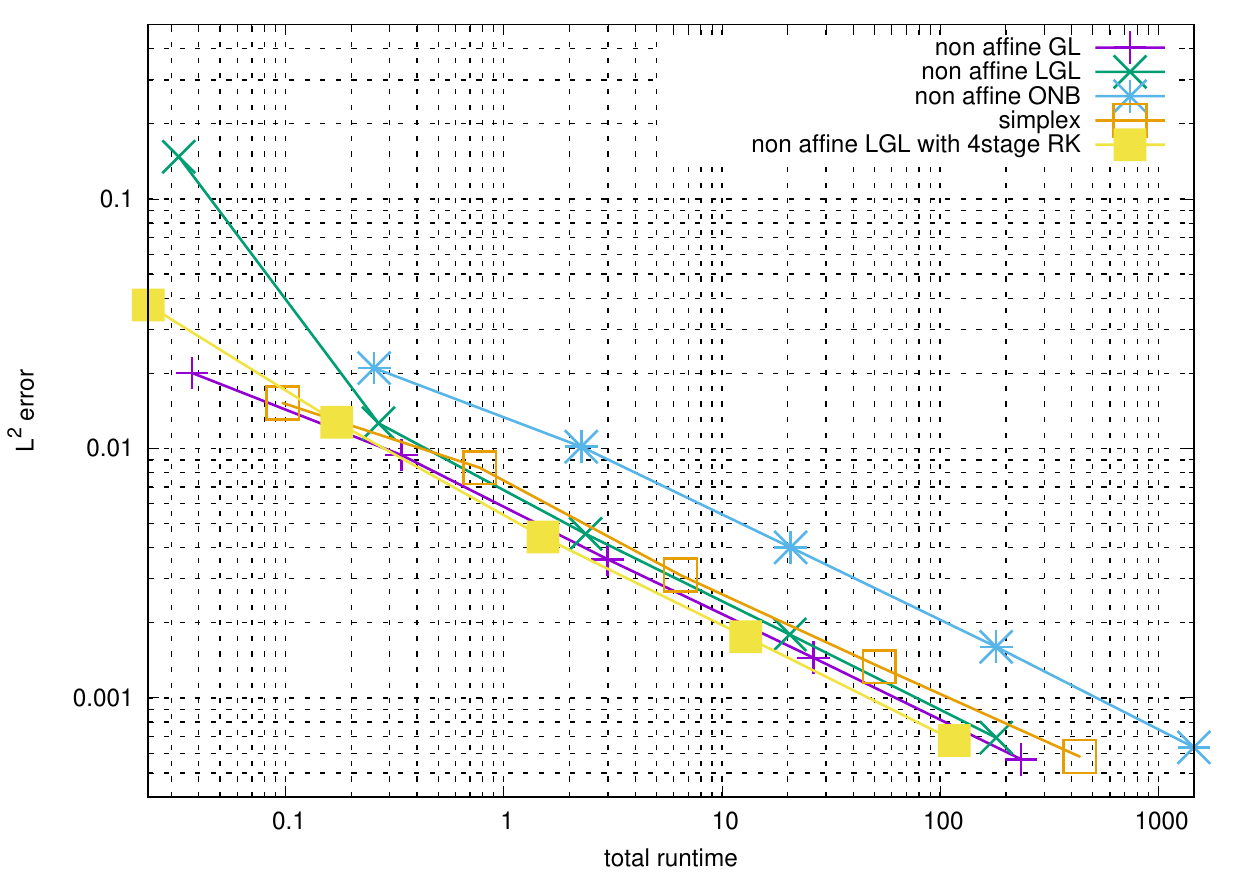}
\caption{Two rarefaction wave problem simulated from $t=0.05$ to $0.12$
on a sequence of non-affine cube and on simplicial grids
with different representation for the discrete space with polynomial order
$4$.}
\label{fig:2vwsmooth}
\end{figure}

\begin{figure}
\includegraphics[width=0.475\textwidth]{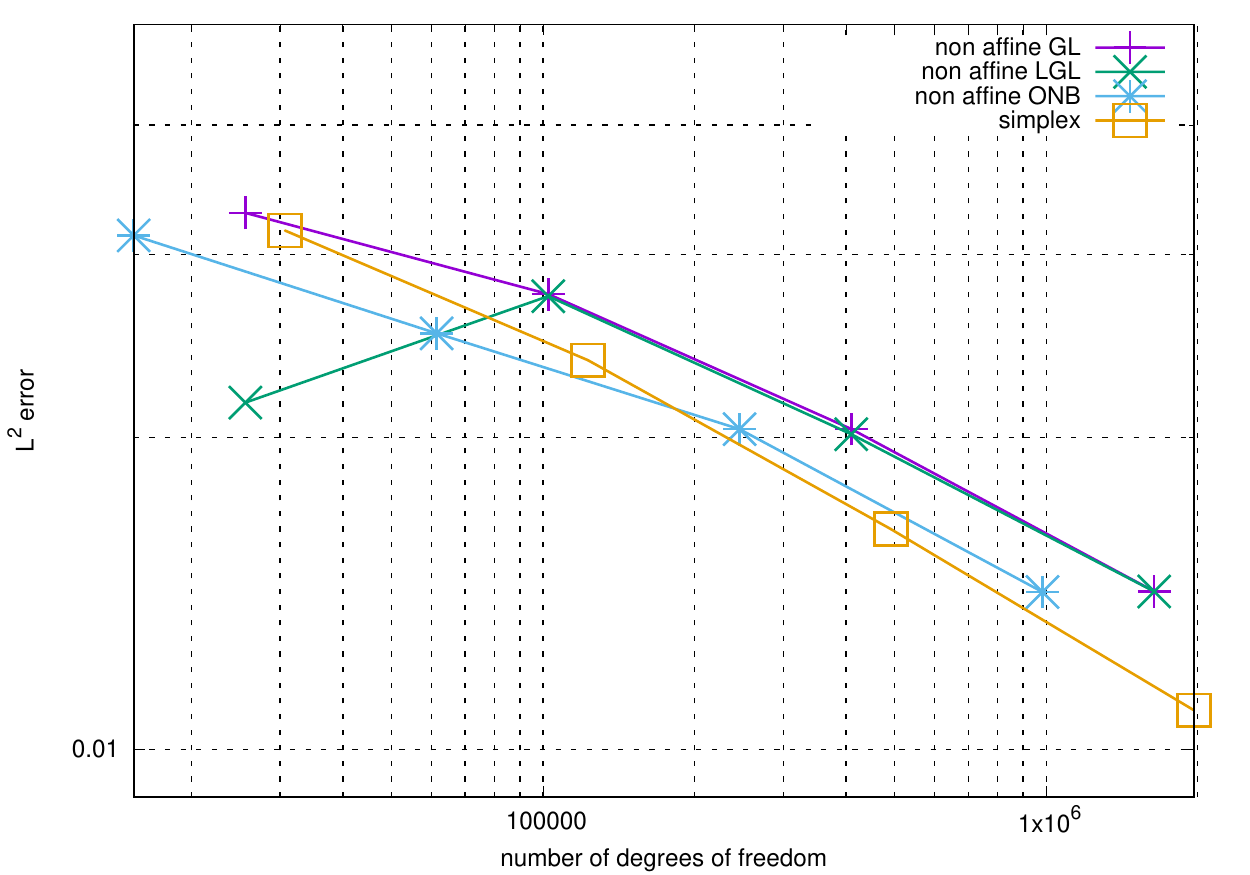}
\includegraphics[width=0.475\textwidth]{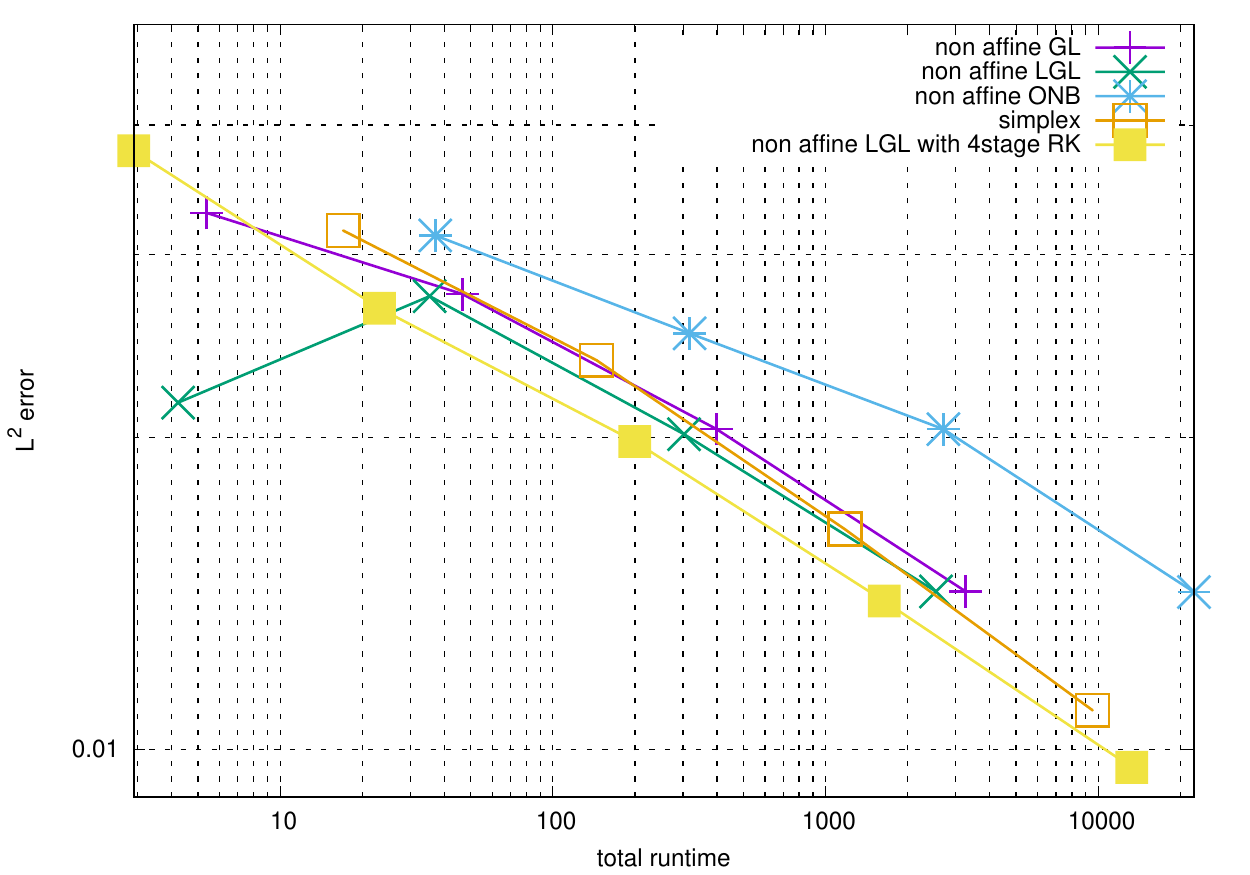}
\caption{Sod's Riemann problem simulated from $t=0$ to $0.2$
on a sequence of non-affine cube and on simplicial grids
with different representation for the discrete space with polynomial order
$4$.}
\label{fig:sod}
\end{figure}

In Fig.~\ref{fig:2vwsmooth}, \ref{fig:sod} we compare $L^2$ errors on a sequence of non affine
cube grids (split into 2 simplices for the simplicial simulation)
using different sets of basis functions.
We show both the error in the $L^2$ vs. the number of degrees of freedom (left) and vs. the
runtime (right). The right plot also includes results from a simulation with the
LGL method and the SSP3 time stepper implemented in Python as discussed previously.
The results of this simulation are not included on the left since they are
broadly in line with the LGL simulation using the 3 stage RK method
available in \dunefemdg.

Summarizing the results from both Fig. \ref{fig:2vwsmooth}, \ref{fig:sod}
it seems clear that the simulation on the simplicial grid produces a
slightly better error on the same grid but requires twice as many
degrees of freedom so that it is less efficient compared to the LGL or GL
simulations. Also as expected, the runtime with the ONB basis on the cube
grids is significantly larger due to the
additional cost of computing the inverse mass matrix on each element of the
grid as discussed above. On an affine cube grid the runtime is comparable
to the GL scheme on the same grid but this is not shown here.
Finally, due to the larger effective CFL constant
the time stepper implemented in Python is more efficient then the 3 stage
method of the same order available in \dunefemdg.

\begin{figure}
\includegraphics[width=0.475\textwidth]{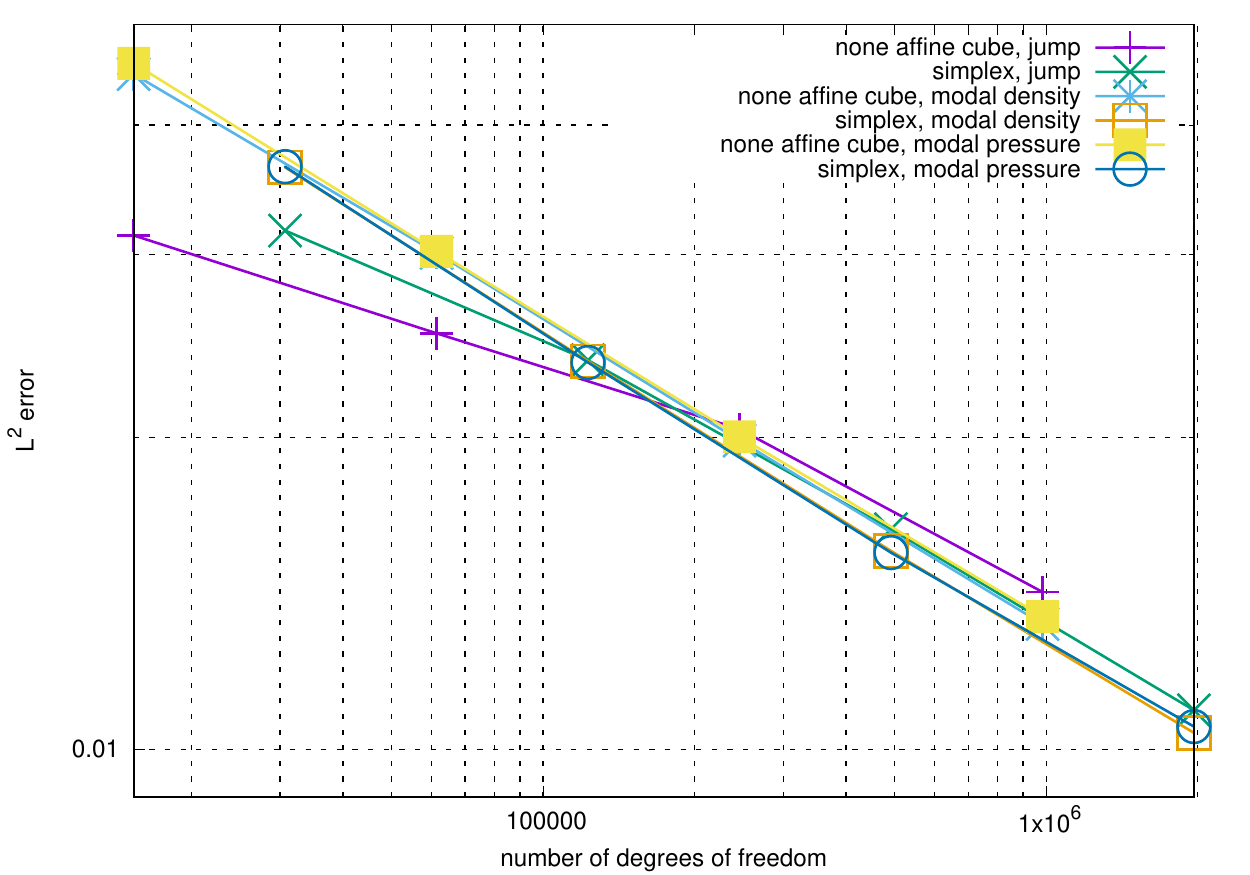}
\includegraphics[width=0.475\textwidth]{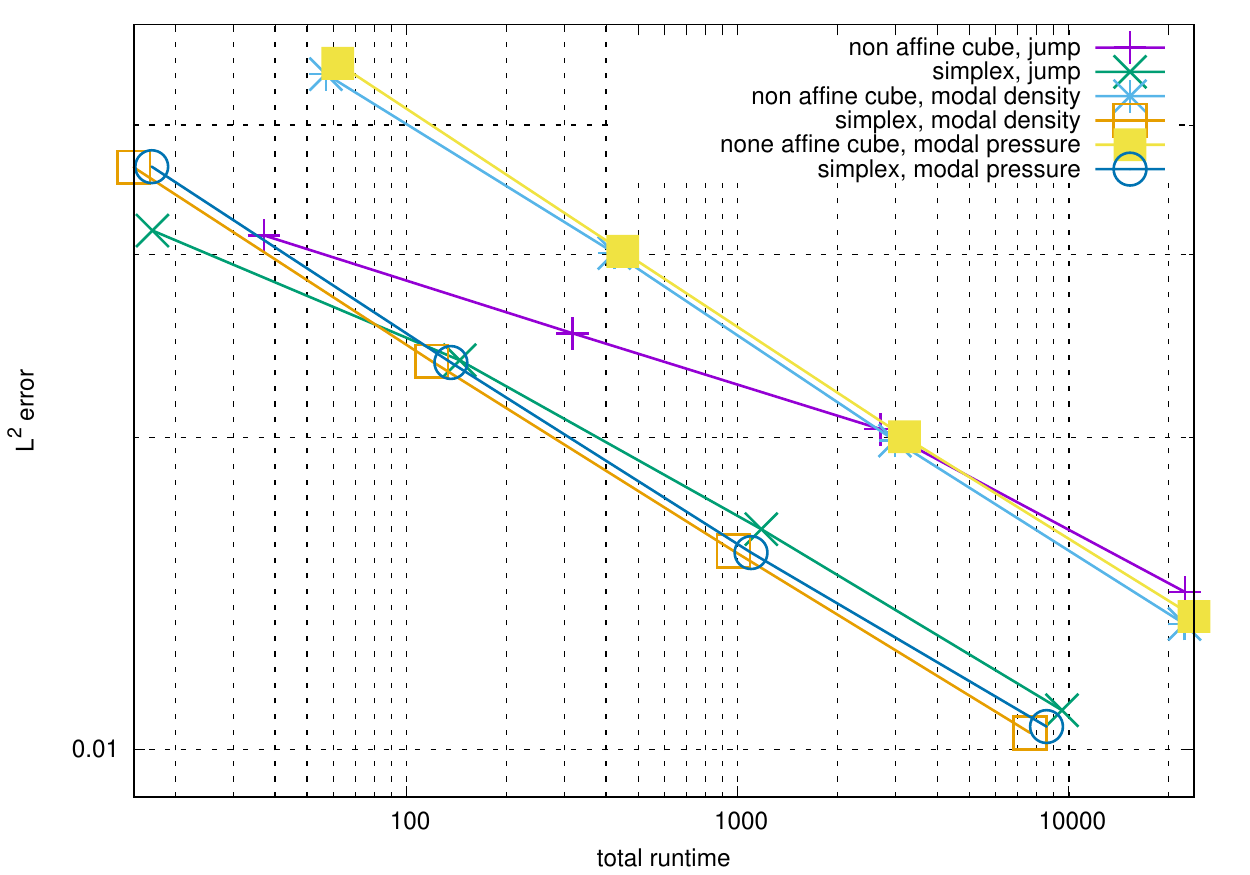}
\caption{Sod's Riemann problem simulated from $t=0$ to $0.2$
using different troubled cell indicators.
The simulations were performed on a sequence of grids consisting
of non affine cubes (split into two triangles for simplicial simulations)
using piecewise polynomials of order $4$ with the orthonormal basis.
Shown are the errors measured in the
$L^2$ norm versus the number of degrees of freedom (left) and the runtime
(right).}
\label{fig:sodtci}
\end{figure}

In Fig. \ref{fig:sodtci} we carry out the same experiment but this time
using the orthonormal basis but varying the troubled cell indicator. While
all method seem to broadly converge in a similar fashion when the grids are
refined, it is clear that the main difference in the efficiency is again
based on question if the grid elements are affine mappings of the reference
element or not. For a given grid structure the results indicate that using
the modal indicator based on the density is the most efficient for the
given test case.





\subsection{Reactive advection diffusion problem}
\label{sec:radproblem}
We conclude with an example demonstrating the flexibility of the framework
to combine different components of the \dunefem package to construct a scheme for a more complex problem.
As a simple example we use a chemical reaction type problem with linear
advection and diffusion where the velocity field is given by discretizing
the solution to an elliptic problem in a continuous Lagrange space. As
mentioned this is still a simple problem but can be seen as a template for
coupled problems, e.g., transport in porous media setting or where the flow is given by 
solving incompressible Navier Stokes equations.

Let us first compute the velocity given as the curl of the solution to a
scalar elliptic problem:
\begin{python}[Computing a velocity field given as curl of solution to Laplace problem]
streamSpace = lagrange(gridView, order=order)
Psi  = streamSpace.interpolate(0,name="streamFunction")
u,v  = TrialFunction(streamSpace), TestFunction(streamSpace)
x    = SpatialCoordinate(streamSpace)
form = ( inner(grad(u),grad(v)) - 5*sin(x[0])*sin(x[1]) * v ) * dx
streamScheme = galerkin([form == 0, DirichletBC(streamSpace,0) ])
streamScheme.solve(target=Psi)
transportVelocity = as_vector([-Psi.dx(1),Psi.dx(0)])
\end{python}
We use this velocity field to evolve three chemical components reacting
through some non-linear reaction term and include some small linear
diffusion:
\begin{python}[Model class for chemical reaction problem]
from ufl import *
from dune.ufl import DirichletBC
from dune.fem.space import lagrange
from dune.fem.scheme import galerkin

class Model:
    dimRange = 3
    # source term (treated explicitly)
    def S_e(t,x,U,DU):
        # reaction term
        r = 10*as_vector([U[0]*U[1], U[0]*U[1], -2*U[0]*U[1]])
        # source for component one and two
        P1 = as_vector([0.2*pi,0.2*pi]) # midpoint of first source
        P2 = as_vector([1.8*pi,1.8*pi]) # midpoint of second source
        f1 = conditional(dot(x-P1,x-P1) < 0.2, 1, 0)
        f2 = conditional(dot(x-P2,x-P2) < 0.2, 1, 0)
        f  = conditional(t<5, as_vector([f1,f2,0]), as_vector([0,0,0]))
        return f - r
    # diffusion term
    def F_v(t,x,U,DU):
        return 0.02*DU
    # advection term
    def F_c(t,x,U):
        return as_matrix([ [*(Model.velocity(t,x,U)*u)] for u in U ])
    def maxWaveSpeed(t,x,U,n):
        return abs(dot(Model.velocity(t,x,U),n))
    # dirichlet boundary conditions
    boundary = {range(1,5): as_vector([0,0,0])}
    # initial conditions
    U0 = [0,0,0]
    endTime = 10
\end{python}
Note that the source term includes both the chemical reaction and a source
for the first two components. The third component is generated by the first
two interacting.

\begin{figure}
  \centering
  \includegraphics[width=0.8\textwidth]{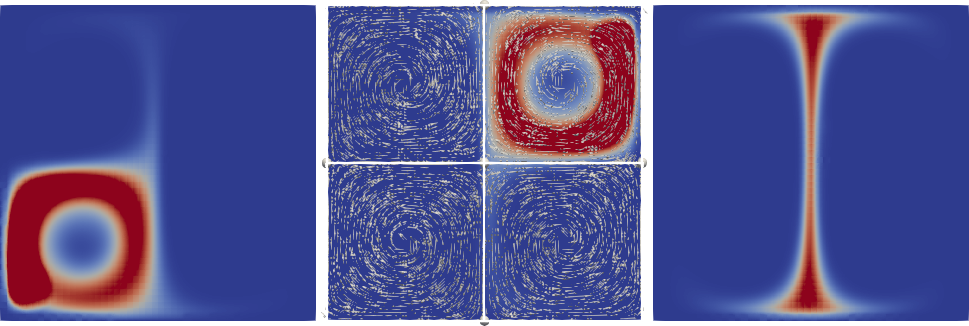} \\
  \includegraphics[width=0.8\textwidth]{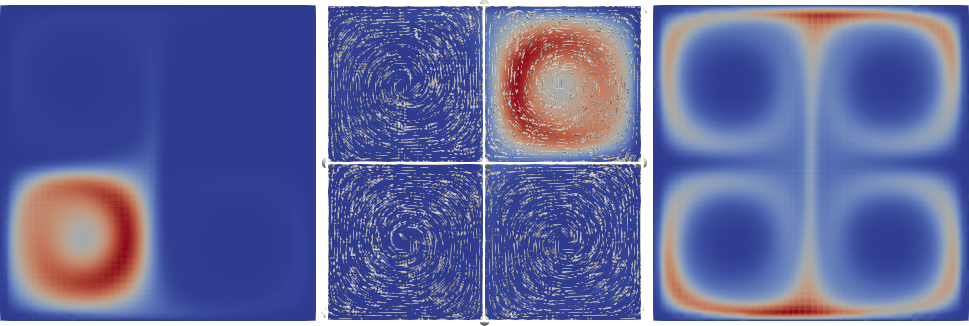}
  \caption{The three components of the chemical reaction system (left to right)
  at $t=4.5$ (top row) and $t=10$ (bottom row). Velocity field included in
  middle figure.}
  \label{fig:chemical}
\end{figure}

Due to the diffusion we do not need any stabilization of the form used so far.
However, in this case a reasonable assumption is that all components remain
positive throughout the simulation, so the physicality check described
above is still a useful feature. We can combine this with the scaling
limiter already described for the advection problem. To this end we need to
add bounds to the model and change to the construction call for the operator:
\begin{python}[Using scaling limiter]
Model.lowerBound = [0,0,0]
operator = femDGOperator(Model, space, limiter="scaling")
\end{python}
Note that by default the stepper switches to a IMEX Runge-Kutta scheme if
the \pyth{Model} class contains both an advective and a diffusive flux.
This behavior can be changed by using the \pyth{rkType} parameter in the
constructor call for the stepper.
A final remark concerning boundary conditions: here we use simple Dirichlet
boundary conditions which are then used for both the diffusive and advective
fluxes on the boundary as outside cell value. 
We saw in a previous example that we can also prescribe the advective flux
on the boundary directly. In that example we used
\begin{python}[Boundary conditions]
Model.boundary[3] = lambda t,x,u,n: noFlowFlux(u,n)
\end{python}
If this type of boundary conditions is used we also need to prescribe
the flux for the diffusion term, so for an advection-diffusion problem we
pass in a pair of fluxes at the boundary, e.g.,
\begin{python}[Flux boundary conditions for advection-diffusion problem]
Model.boundary[3] = [ lambda t,x,U,n:    inner( Model.Fc(t,x,Ubnd), n ),
                      lambda t,x,U,DU,n: inner( Model.Fv(t,x,Ubnd,DUbnd), n ) ]
\end{python}
Fig. \ref{fig:chemical} shows the results for the chemical reaction
problem.

\section{Efficiency of Python based auto-generated models} \label{sec:codegen}

While Python is easy to use, its flexibility can
lead to some deficiencies when it comes to performance.
In \dune \cite{dune-python,dunefempy} a just-in-time compilation concept is used to
create Python modules based on the static C++ type of every object used, i.e. 
the models described in the previous section are translated into C++ code based on the UFL descriptions in
the various model methods which is then compiled and loaded as Python modules. 
This way we avoid virtualization of the \dune interfaces and consequently
one would expect very little performance impact as long as calls between
Python and C++ are only done for long running methods. To verify this,
we compare the performance of the approach shown here with
the previously hand-coded pure C++ version described in \cite{dunefemdg:17}.

As a test example we choose a standard Riemann problem (Sod,  $T=0.1$) for the Euler
equations solved on a series of different grid resolutions using forth order basis functions, 
the default limiter, and explicit RK3 time stepping with $CFL=0.4$.
We use two different grid implementation, a dedicated Cartesian grid
(SPGrid) and a fully unstructured grid (ALUGrid).
The results are shown in Table~\ref{tab:performance} and should be taken with a
grain of salt since these only present a preliminary comparison, as 
such measurements heavily depend on the hardware, compiler
and compiler flags as well as selected mathematical problem. 
The results presented here were produced on an Intel CPU i7-9750H @ 2.60GHz
using \pyth{g++-9.3} and the following compiler flags: 
\begin{pythonline}
 -O3 -DNDEBUG -Wfatal-errors -march=native -fomit-frame-pointer -ftree-vectorize 
  -fexpensive-optimizations --param large-unit-insns=500 
  --param inline-unit-growth=500  --param max-inline-insns-single=500  
  --param large-function-growth=500  --param large-function-insns=500
\end{pythonline}

\begin{table}
\renewcommand{\arraystretch}{1.2}
  \caption{Runtime comparison of the C++ and the Python code for a simple test
  example solving the Euler equations in 2d with explicit time stepping and
  forth order polynomials using $1$ and $4$ thread(s).}
  \label{tab:performance}
\nopagebreak
\begin{tabular}{lr}
    \multicolumn{2}{c}{4th order polynomials and 1 thread.} \\
\begin{minipage}{0.48\textwidth}
  \begin{center}
  SPGrid\\
  \begin{tabular}{l|ccc}
     code%
$\backslash$ \#el &   $1024$    &    $4096$    &   $16384$  \\  \hline\hline
    C++           &   15.5      &    128.6     &    1045.4  \\ 
    C++(UFL)      &   14.8      &    123.7     &    1013.2  \\
    Python        &   14.8      &    123.9     &    1008.6  \\
  \end{tabular}
  \end{center}
\end{minipage} &
\begin{minipage}{0.48\textwidth}
  \begin{center}
  ALUGrid \\
  \begin{tabular}{l|ccc}
     code%
  $\backslash$ \#el  &  $1024$   &  $4096$   &  $16384$  \\  \hline\hline
    C++              &   17.3    &   145.6   &   1203.5  \\
    C++(UFL)         &   17. 3   &   145.0   &   1199.3  \\
    Python           &   19.5    &   163.3   &   1341.7  \\   
  \end{tabular}
  \end{center}
\end{minipage}
\end{tabular}
\label{tab:performance41}

\vskip 12pt 
\begin{tabular}{lr}
    \multicolumn{2}{c}{4th order polynomials and 4 threads.} \\
\begin{minipage}{0.48\textwidth}
  \begin{center}
  SPGrid\\
  \begin{tabular}{l|ccc}
     code%
$\backslash$ \#el &   $1024$    &    $4096$    &   $16384$  \\  \hline\hline
    C++           &    4.43     &     35.7     &    291.2  \\ 
    C++(UFL)      &    4.18     &     34.4     &    282.0  \\
    Python        &    4.19     &     34.1     &    280.9  \\
  \end{tabular}
  \end{center}
\end{minipage} &
\begin{minipage}{0.48\textwidth}
  \begin{center}
  ALUGrid \\
  \begin{tabular}{l|ccc}
     code%
  $\backslash$ \#el  &  $1024$   &  $4096$   &  $16384$  \\  \hline\hline
    C++              &   4.79    &   39.9    &   331.3   \\
    C++(UFL)         &   4.73    &   39.9    &   331.5   \\
    Python           &   5.43    &   44.2    &   365.7   \\
  \end{tabular}
  \end{center}
\end{minipage}
\end{tabular}
\label{tab:performance44}
\end{table}

With our setting we observe that for the Cartesian grid (SPGrid) using the Python front-end
leads to no performance loss.
For the unstructured grid (ALUGrid) we observe a performance decrease of about 10\%.
This can be explained with the fact that for SPGrid all code can be
inlined in the just-in-time compiled Python module which uses the same compiler
optimization flags as the normal C++ code. For ALUGrid, where a
shared library exists, this is not so straight forward. In the future we will
experiment with link time optimization and try to reduce implementation of
small code snippets in the ALUGrid library.

\section{Conclusion and outlook}
\label{sec:conclusion}

In this paper we presented a comprehensive framework for the 
Discontinuous Galerkin (DG) method
with an easy to use Python interface for model description and solver
configuration. The framework covers different variants of existing DG methods 
and can be easily extended to include improvements and new
development of the methodology. 

Although the \dunefemdg framework serves as a good starting point for
high-performance DG solvers, the aim of this work is to demonstrate how the
Python interface simplifies the simulation of a wide range of evolution
equations as well as enabling the rapid prototyping of new methods, their
testing and their comparison with other methods. 
For example, in almost all the tests presented here, computations were done 
using the same setting for the indicator and tolerance for adaptivity and 
the troubled cell detection, demonstrating the robustness of the default
setup. Of course the default setup will only be a suitable starting point
and so most of the building blocks of the discretization can be
straightforwardly replaced from within the Python script,
often without requiring any or not much C++ knowledge.
Also we would like to point out the implemented DG solver 
can be directly used as higher (2nd) order Finite Volume (FV) scheme 
by replacing the DG space with a FV space consisting of piecewise constant polynomials.

The Python interfaces are not yet fully reflecting the possibilities on the C++
side, for example, the reconstruction in primary variables for the Euler equations
is possible, but not yet available on the Python side. 
Another missing feature 
is that it is currently not possible on the Python side to easily 
exchange the average operators in the discretization of the diffusive terms 
as, for example, needed in some applications \cite{twophase:18}.

Another missing feature in Python is the assembly of Jacobian matrices 
during the nonlinear solve. This could be desirable for some applications or 
to test different existing solvers.
Although the feature is available in the C++ code it needs a few code
alterations in the underlying infrastructure package. This would also make
it straightforward to then use other solver packages available in Python,
e.g., \pyth{scipy}.
For the compressible applications, which were the focus of this work,
it is much more feasible to 
work with the Jacobian free nonlinear solvers and there robust and efficient
preconditioning techniques are still a very active research topic. 
We are currently focusing on including ideas presented in 
\cite{birken:19} based on subcell finite volume multigrid preconditioning.

The extension of the DGSEM methods to simplicial grids 
(see \cite{dgsem-review:20} for an overview) could be implemented 
with a few minor modifications. Based on earlier work \cite{DednerNolte:12} 
the necessary basis function implementations for arbitrary quadrature points are
available but need to be integrated into the \dunefem and \dunefemdg framework.

While \dunefem provides a number of Runge-Kutta methods,
we have shown here that it is straightforward to add other time stepping algorithms
on the Python side, a feature which should also allow to use other packages
which provide bindings for Python, such as the Assimulo package \cite{Andersson2015}.
For IMEX schemes the splitting is at the
moment still a bit restricted focusing on the important case of implicitly
treating the diffusion (and part of the source term) while using an
explicit method for the advection. In some applications other types of
splitting (e.g. \emph{horizontal explicit vertical implicit} (HEVI) methods
used in meteorology) are of interest and will be made available in future releases.

\section*{Conflict of interest}
On behalf of all authors, the corresponding author states that there is no conflict of interest.

\bibliographystyle{spmpsci}
\bibliography{bibliography}

\begin{thebibliography}{10}
\providecommand{\url}[1]{{#1}}
\providecommand{\urlprefix}{URL }
\expandafter\ifx\csname urlstyle\endcsname\relax
  \providecommand{\doi}[1]{DOI~\discretionary{}{}{}#1}\else
  \providecommand{\doi}{DOI~\discretionary{}{}{}\begingroup
  \urlstyle{rm}\Url}\fi

\bibitem{UFL}
Aln{\ae}s, M.S., Logg, A., {\O}lgaard, K.B., Rognes, M.E., Wells, G.N.:
  {Unified Form Language: {A} domain-specific language for weak formulations of
  partial differential equations}.
\newblock CoRR \textbf{abs/1211.4047} (2012).
\newblock \urlprefix\url{http://arxiv.org/abs/1211.4047}

\bibitem{Andersson2015}
Andersson, C., F\"uhrer, C., {\AA}kesson, J.: Assimulo: A unified framework for
  \{ODE\} solvers.
\newblock Mathematics and Computers in Simulation \textbf{116}(0), 26 -- 43
  (2015).
\newblock \doi{10.1016/j.matcom.2015.04.007}

\bibitem{petsc-user-ref}
Balay, S., Abhyankar, S., Adams, M.F., Brown, J., Brune, P., Buschelman, K.,
  Dalcin, L., Dener, A., Eijkhout, V., Gropp, W.D., Karpeyev, D., Kaushik, D.,
  Knepley, M.G., May, D.A., McInnes, L.C., Mills, R.T., Munson, T., Rupp, K.,
  Sanan, P., Smith, B.F., Zampini, S., Zhang, H., Zhang, H.: {PETS}c users
  manual.
\newblock Tech. Rep. ANL-95/11 - Revision 3.14, Argonne National Laboratory
  (2020).
\newblock \urlprefix\url{https://www.mcs.anl.gov/petsc}

\bibitem{BangerthHartmannKanschat2007}
Bangerth, W., Hartmann, R., Kanschat, G.: {deal.II} -- a general purpose object
  oriented finite element library.
\newblock ACM Trans. Math. Softw. \textbf{33}(4), 24/1--24/27 (2007).
\newblock \urlprefix\url{http://dealii.org/}

\bibitem{dunepaperII:08}
Bastian, P., Blatt, M., Dedner, A., Engwer, C., Kl{\"o}fkorn, R., Kornhuber,
  R., Ohlberger, M., Sander, O.: {A Generic Grid Interface for Parallel and
  Adaptive Scientific Computing. Part {II}: Implementation and Tests in
  {DUNE}}.
\newblock Computing \textbf{82}(2--3), 121--138 (2008).
\newblock \doi{10.1007/s00607-008-0004-9}

\bibitem{dunereview:20}
Bastian, P., Blatt, M., Dedner, M., Dreier, N.A., Engwer Ch.~Fritze, R.,
  Gr{\"a}ser, C., Gr{\"u}ninger, C., Kempf, D., Kl{\"o}fkorn, R., Ohlberger,
  M., Sander, O.: {The Dune framework: Basic concepts and recent developments}.
\newblock {CAMWA}  (2020).
\newblock \doi{10.1016/j.camwa.2020.06.007}

\bibitem{birken:19}
Birken, P., Gassner, G.J., Versbach, L.M.: Subcell finite volume multigrid
  preconditioning for high-order discontinuous galerkin methods.
\newblock International Journal of Computational Fluid Dynamics \textbf{33}(9),
  353--361 (2019).
\newblock \doi{10.1080/10618562.2019.1667983}

\bibitem{dunecosmo:12}
Brdar, S., Baldauf, M., Dedner, A., Kl{\"o}fkorn, R.: {Comparison of dynamical
  cores for NWP models: comparison of COSMO and {DUNE}}.
\newblock Theoretical and Computational Fluid Dynamics \textbf{27}(3-4),
  453--472 (2013).
\newblock \doi{10.1007/s00162-012-0264-z}

\bibitem{cdg2:12}
Brdar, S., Dedner, A., Kl{\"o}fkorn, R.: {Compact and stable Discontinuous
  Galerkin methods for convection-diffusion problems.}
\newblock SIAM J. Sci. Comput. \textbf{34}(1), 263--282 (2012).
\newblock \doi{10.1137/100817528}

\bibitem{Chen:LPreconstruction}
Chen, L., Li, R.: {An Integrated Linear Reconstruction for Finite Volume scheme
  on Unstructured Grids}.
\newblock J. Sci. Comput. \textbf{68}, 1172--1197 (2016).
\newblock \doi{10.1007/s10915-016-0173-1}

\bibitem{dgsem-review:20}
Chen, T., Shu, C.W.: {Review Article: Review of Entropy Stable Discontinuous
  Galerkin Methods for Systems of Conservation Laws on Unstructured Simplex
  Meshes}.
\newblock CSIAM Transactions on Applied Mathematics \textbf{1}(1), 1--52
  (2020).
\newblock \doi{10.4208/csiam-am.2020-0003}

\bibitem{cheng:13}
Cheng, Y., Li, F., Qiu, J., Xu, L.: {Positivity-preserving DG and central DG
  methods for ideal MHD equations}.
\newblock Journal of Computational Physics \textbf{238}, 255 -- 280 (2013).
\newblock \doi{10.1016/j.jcp.2012.12.019}

\bibitem{CSreview}
Cockburn, B., Shu, C.W.: {R}unge-{K}utta {D}iscontinuous {G}alerkin methods for
  convection-dominated problems.
\newblock J. Sci. Comput. \textbf{16}(3), 173--261 (2001)

\bibitem{dunefemdg:17}
Dedner, A., Girke, S., Kl{\"o}fkorn, R., Malkmus, T.: {The DUNE-FEM-DG module}.
\newblock ANS \textbf{5}(1) (2017).
\newblock \doi{10.11588/ans.2017.1.28602}

\bibitem{twophase:18}
Dedner, A., Kane, B., Kl{\"o}fkorn, R., Nolte, M.: {Python framework for
  hp-adaptive discontinuous Galerkin methods for two-phase flow in porous
  media}.
\newblock AMM \textbf{67}, 179 -- 200 (2019).
\newblock \doi{10.1016/j.apm.2018.10.013}

\bibitem{dunefempy}
Dedner, A., Kloefkorn, R., Nolte, M.: Python bindings for the dune-fem module
  (2020).
\newblock \doi{10.5281/zenodo.3706994}

\bibitem{limiter:11}
Dedner, A., Kl{\"o}fkorn, R.: {A Generic Stabilization Approach for Higher
  Order Discontinuous Galerkin Methods for Convection Dominated Problems}.
\newblock J. Sci. Comput. \textbf{47}(3), 365--388 (2011).
\newblock \doi{10.1007/s10915-010-9448-0}

\bibitem{dunefemdgrepo}
Dedner, A., Kl\"ofkorn, R.: {The Dune-Fem-DG Module}.
\newblock \url{https://gitlab.dune-project.org/dune-fem/dune-fem-dg} (2019)

\bibitem{fvca9:20}
Dedner, A., Kl{\"o}fkorn, R.: {A Python Framework for Solving
  Advection-Diffusion Problems}.
\newblock In: R.~Kl{\"o}fkorn, E.~Keilegavlen, F.A. Radu, J.~Fuhrmann (eds.)
  Finite Volumes for Complex Applications IX - Methods, Theoretical Aspects,
  Examples, pp. 695--703. Springer International Publishing, Cham (2020).
\newblock \doi{10.1007\%2F978-3-030-43651-3_66}

\bibitem{dune:Fem}
Dedner, A., Kl{\"o}fkorn, R., Nolte, M., Ohlberger, M.: {A Generic Interface
  for Parallel and Adaptive Scientific Computing: abstraction Principles and
  the DUNE-FEM Module}.
\newblock Computing \textbf{90}(3--4), 165--196 (2010).
\newblock \doi{10.1007/s00607-010-0110-3}

\bibitem{DMO:07}
Dedner, A., Makridakis, C., Ohlberger, M.: {Error control for a class of
  Runge-Kutta discontinuous Galerkin methods for nonlinear conservation laws.}
\newblock SIAM J. Numer. Anal. \textbf{45}(2), 514--538 (2007).
\newblock \doi{10.1137/050624248}

\bibitem{DednerNolte:12}
Dedner, A., Nolte, M.: {Construction of Local Finite Element Spaces Using the
  Generic Reference Elements}.
\newblock In: A.~Dedner, B.~Flemisch, R.~Kl{\"o}fkorn (eds.) Advances in DUNE,
  pp. 3--16. Springer Berlin Heidelberg, Berlin, Heidelberg (2012).
\newblock \doi{10.1007/978-3-642-28589-9_1}

\bibitem{dune-python}
Dedner, A., Nolte, M.: {The Dune-Python Module}.
\newblock CoRR \textbf{abs/1807.05252} (2018).
\newblock \urlprefix\url{http://arxiv.org/abs/1807.05252}

\bibitem{Discacciati:20}
Discacciati, N., Hesthaven, J.S., Ray, D.: {Controlling oscillations in
  high-order Discontinuous Galerkin schemes using artificial viscosity tuned by
  neural networks}.
\newblock Journal of Computational Physics \textbf{409}, 109304 (2020).
\newblock \doi{10.1016/j.jcp.2020.109304}

\bibitem{Dolejsi:03}
Dolejší, V., Feistauer, M., Schwab, C.: {On some aspects of the discontinuous
  Galerkin finite element method for conservation laws}.
\newblock Mathematics and Computers in Simulation \textbf{61}(3), 333 -- 346
  (2003).
\newblock \doi{10.1016/S0378-4754(02)00087-3}

\bibitem{Dumbser:08}
Dumbser, M., Balsara, D.S., Toro, E.F., Munz, C.D.: {A unified framework for
  the construction of one-step finite volume and discontinuous Galerkin schemes
  on unstructured meshes}.
\newblock Journal of Computational Physics \textbf{227}(18), 8209 -- 8253
  (2008).
\newblock \doi{10.1016/j.jcp.2008.05.025}

\bibitem{feelpp}
\feelconsortium: The Feel++ Book (2015).
\newblock \urlprefix\url{https://www.gitbook.com/book/feelpp/feelpp-book}

\bibitem{feist:08}
Feistauer, M., Ku\v{c}era, V.: {A new technique for the numerical solution of
  the compressible Euler equations with arbitrary Mach numbers.}
\newblock {Benzoni-Gavage, Sylvie (ed.) et al., Hyperbolic problems. Theory,
  numerics and applications. Proceedings of the 11th international conference
  on hyperbolic problems, Ecole Normale Sup\'erieure, Lyon, France, July
  17--21, 2006. Berlin: Springer. 523-531 (2008).} (2008)

\bibitem{shu:01}
Gottlieb, S., Shu, C.W., Tadmor, E.: {Strong Stability-Preserving High-Order
  Time Discretization Methods}.
\newblock SIAM Rev. \textbf{43}(1), 89--112 (2001).
\newblock \doi{10.1137/S003614450036757X}

\bibitem{Guermond:11}
Guermond, J.L., Pasquetti, R., Popov, B.: Entropy viscosity method for
  nonlinear conservation laws.
\newblock Journal of Computational Physics \textbf{230}(11), 4248 -- 4267
  (2011).
\newblock \doi{10.1016/j.jcp.2010.11.043}.
\newblock Special issue High Order Methods for CFD Problems

\bibitem{flexi:12}
Hindenlang, F., Gassner, G.J., Altmann, C., Beck, A., Staudenmaier, M., Munz,
  C.D.: {Explicit discontinuous Galerkin methods for unsteady problems}.
\newblock Computers \& Fluids \textbf{61}, 86 -- 93 (2012).
\newblock \doi{10.1016/j.compfluid.2012.03.006}

\bibitem{exastencils:20}
H{\"o}nig, J., Koch, M., R{\"u}de, U., Engwer, C., K{\"o}stler, H.: {Unified}
  generation of {DG}-kernels for different {HPC} frameworks.
\newblock In: I.~Foster, G.R. Joubert, L.~Kucera, W.E. Nagel, F.~Peters (eds.)
  Advances in Parallel Computing, vol.~36, pp. 376--386. IOS Press BV (2020).
\newblock \doi{10.3233/APC200062}

\bibitem{Houston:18}
Houston, P., Sime, N.: {Automatic Symbolic Computation for Discontinuous
  Galerkin Finite Element Methods}.
\newblock SIAM Journal on Scientific Computing \textbf{40}(3), C327--C357
  (2018).
\newblock \doi{10.1137/17M1129751}

\bibitem{ka.sh:05}
Karniadakis, G., Sherwin, S.: Spectral/hp element methods for computational
  fluid dynamics.
\newblock Oxford University Press (2005).
\newblock \urlprefix\url{http://www.nektar.info/}

\bibitem{Ketcheson:08}
Ketcheson, D.I.: {Highly Efficient Strong Stability-Preserving Runge-Kutta
  Methods with Low-Storage Implementations}.
\newblock SIAM Journal on Scientific Computing \textbf{30}(4), 2113--2136
  (2008).
\newblock \doi{10.1137/07070485X}

\bibitem{rivlimiter:06}
Klieber, W., Rivi\`ere, B.: {Adaptive simulations of two-phase flow by
  discontinuous Galerkin methods.}
\newblock Comput. Methods Appl. Mech. Eng. \textbf{196}(1-3), 404--419 (2006).
\newblock \doi{10.1016/j.cma.2006.05.007}

\bibitem{Kloeckner}
{Kl\"ockner, A.}, {Warburton, T.}, {Hesthaven, J. S.}: Viscous shock capturing
  in a time-explicit discontinuous galerkin method.
\newblock Math. Model. Nat. Phenom. \textbf{6}(3), 57--83 (2011).
\newblock \doi{10.1051/mmnp/20116303}

\bibitem{dgimpl:12}
Kl{\"o}fkorn, R.: {Efficient Matrix-Free Implementation of Discontinuous
  Galerkin Methods for Compressible Flow Problems}.
\newblock In: A.H. et~al. (ed.) Proceedings of the ALGORITMY 2012, pp. 11--21
  (2012)

\bibitem{reconpoly:17}
Kl{\"o}fkorn, R., Kvashchuk, A., Nolte, M.: {Comparison of linear
  reconstructions for second-order finite volume schemes on polyhedral grids}.
\newblock Computational Geosciences \textbf{21}(5), 909--919 (2017).
\newblock \doi{10.1007/s10596-017-9658-8}

\bibitem{knoll:04}
Knoll, D.A., Keyes, D.E.: {Jacobian-free Newton-Krylov methods: a survey of
  approaches and applications.}
\newblock J. Comput. Phys. \textbf{193}(2), 357--397 (2004)

\bibitem{kopriva:10}
Kopriva, D.A., Gassner, G.: On the quadrature and weak form choices
  in collocation type discontinuous galerkin spectral element methods.
\newblock Journal of Scientific Computing \textbf{44}, 136--155 (2010).
\newblock \doi{10.1007/s10915-010-9372-3}

\bibitem{kopriva:02}
Kopriva, D.A., Woodruff, S.L., Hussaini, M.Y.: {Computation of electromagnetic
  scattering with a non-conforming discontinuous spectral element method}.
\newblock International Journal for Numerical Methods in Engineering
  \textbf{53}(1), 105--122 (2002).
\newblock \doi{10.1002/nme.394}

\bibitem{Krivodonova2004}
Krivodonova, L., Xin, J., Remacle, J.F., Chevaugeon, N., Flaherty, J.E.: {Shock
  detection and limiting with discontinuous Galerkin methods for hyperbolic
  conservation laws.}
\newblock Appl. Numer. Math. \textbf{48}(3-4), 323--338 (2004).
\newblock \doi{10.1016/j.apnum.2003.11.002}

\bibitem{fenics}
Logg, A., Mardal, K.A., Wells, G.: Automated Solution of Differential Equations
  by the Finite Element Method: The FEniCS Book.
\newblock Springer Publishing Company, Incorporated (2012)

\bibitem{mandli2016clawpack}
Mandli, K.T., Ahmadia, A.J., Berger, M., Calhoun, D., George, D.L.,
  Hadjimichael, Y., Ketcheson, D.I., Lemoine, G.I., LeVeque, R.J.: Clawpack:
  building an open source ecosystem for solving hyperbolic pdes.
\newblock PeerJ Computer Science \textbf{2}, e68 (2016).
\newblock \doi{10.7717/peerj-cs.68}

\bibitem{May2013}
May, S., Berger, M.: Two-dimensional slope limiters for finite volume schemes
  on non-coordinate-aligned meshes.
\newblock {SIAM Journal on Scientific Computing} \textbf{35}(5), A2163--A2187
  (2013).
\newblock \doi{10.1137/120875624}

\bibitem{persson:06}
Persson, P.O., Peraire, J.: Sub-cell shock capturing for discontinuous galerkin
  methods.
\newblock presented at 44th aiaa aerospace sciences meeting, aiaa-2006-0112,
  reno, nevada, Department of Aeronautics \& Astronautics Massachusetts
  Institute of Technology (2006).
\newblock \doi{10.2514/6.2006-112}

\bibitem{firedrake}
Rathgeber, F., Ham, D.A., Mitchell, L., Lange, M., Luporini, F., {McRae},
  A.T.T., Bercea, G.T., Markall, G.R., Kelly, P.H.J.: {Firedrake: Automating
  the Finite Element Method by Composing Abstractions}.
\newblock ACM Trans. Math. Softw. \textbf{43}(3), 24:1--24:27 (2016).
\newblock \doi{10.1145/2998441}

\bibitem{cosmodg:14}
Schuster, D., Brdar, S., Baldauf, M., Dedner, A., Kl{\"o}fkorn, R., Kr{\"o}ner,
  D.: {On discontinuous Galerkin approach for atmospheric flow in the mesoscale
  with and without moisture}.
\newblock Meteorologische Zeitschrift \textbf{23}(4), 449--464 (2014).
\newblock \doi{10.1127/0941-2948/2014/0565}

\bibitem{Shu-boundpreserving-review:16}
Shu, C.W.: {High order WENO and DG methods for time-dependent
  convection-dominated PDEs: A brief survey of several recent developments}.
\newblock Journal of Computational Physics \textbf{316}, 598 -- 613 (2016).
\newblock \doi{10.1016/j.jcp.2016.04.030}

\bibitem{wallwork:20}
Wallwork, J.G., Barral, N., Kramer, S.C., Ham, D.A., Piggott, M.D.:
  Goal-oriented error estimation and mesh adaptation for shallow water
  modelling.
\newblock SN Applied Sciences \textbf{2} (2020).
\newblock \doi{10.1007/s42452-020-2745-9}

\bibitem{zhang:10}
Zhang, X., Shu, C.W.: {On positivity-preserving high order discontinuous
  Galerkin schemes for compressible Euler equations on rectangular meshes}.
\newblock Journal of Computational Physics \textbf{229}(23), 8918 -- 8934
  (2010).
\newblock \doi{10.1016/j.jcp.2010.08.016}

\end{thebibliography}


\begin{appendix}

\section{Installation}
\label{sec:installation}
The presented software is based on the upcoming \dune release version 2.8. 
Installing the software described in this paper can be done using the Package
Installer for Python (pip). This method of installing 
the software has been tested on different Linux systems and MAC OS Catalina. 
Installations on Windows systems require to make use of the \textit{Windows Subsystem for Linux} and,
for example, Ubuntu as an operating system. 

Prerequisites for the installation are 
\begin{enumerate}
  \item[A] A working compiler suite (C++, C) that
supports C++ standard 17 (i.e. g++ version 8 or later or clang version 10 or
later) 
\item[B] pkg-config 
\item[C] cmake version 3.13.3 or later (newer versions can be installed using pip),
\item[D] git for downloading the software, and 
\item[E] a working Python 3 installation including the virtual environment module
  (venv).
\end{enumerate}

The first step is to create a so called Python virtual environment which will
contain all the installed software and for later removal one only has to remove
the folder containing the virtual environment. 

\begin{enumerate}

  \item Create a Python virtual environment, i.e. 
    \begin{pythonline}
  python3 -m venv venv-femdg  
\end{pythonline}
    This will create a virtual environment in the folder \pyth{venv-femdg}.

  \item Activate the virtual environment by 
\begin{pythonline}
  source venv-femdg/bin/activate 
\end{pythonline}  

  The prompt of the shell will now contain the name of the virtual environment
    directory in parenthesis. The virtual environment can be deactivated by
    closing the shell or invoking the command \pyth{deactivate}. 

  \item To ensure consistent package installation upgrade \pyth{pip} first, i.e. 
\begin{pythonline}
  pip install --upgrade pip
\end{pythonline}  

\item Install Python packages needed by scripts in this paper that are not 
  listed as default dependencies of \dunefemdg, i.e. 
\begin{pythonline}
  pip install matplotlib cmake
\end{pythonline}  
  At this point cmake only needs to be installed if the system version is older
  than 3.13.3.

\item Install \dunefemdg by simply using pip again, i.e.
\begin{pythonline}
  pip install dune-fem-dg 
\end{pythonline}  
This will also install all necessary other packages such as numpy, fenics-ufl, mpi4py and
    various \dune packages.
Depending on the time of the day it might be a good idea to get a cup of coffee now, 
because this will take a few minutes since libraries and Python packages are now actually build. 

\item Finally, the scripts used to produce the results presented in this paper
  can be downloaded by calling 
\begin{pythonline}
  python -m dune.femdg  
\end{pythonline}  
    This will create a folder \pyth{femdg_tutorial} containing various example
    programs. Note that the first run of each script will take relatively long
    since various C++ modules are compiled just-in-time.

\end{enumerate}

The software can also be installed using the standard \dune way of installing
from source. A ready to use and up to date build script is found inside the
dune-fem-dg git repository at \url{https://gitlab.dune-project.org/dune-fem/dune-fem-dg/-/blob/master/scripts/build-dune-fem-dg.sh}. 
This way of installation requires more in-depth knowledge of the \dune build
system and Linux expert knowledge in general.

\section{Interfaces}
\label{sec:extension}

Signature of \pyth{femdgOperator}:
\begin{pythonline}
def femDGOperator(Model, space, limiter="default",
      advectionFlux="default", diffusionScheme = "cdg2", parameters=None)
\end{pythonline}
The \pyth{Model} class contains the description of the mathematical model
to solve and is described below. The \pyth{space} parameter is one of the
available Discontinuous Galerkin spaces available in the \dunefem
framework.
We have \pyth{limiter} equal to \pyth{None,"default","minmod","scaling"},
taking \pyth{advectionFlux} equal to \pyth{"default"} to use
Local-Lax-Friedrichs scheme, for Euler there are some other fluxes
available, this parameter can also be used to pass in a user defined flux
implementation.
The parameter \pyth{diffusionScheme} can be used to change the diffusive
flux. The \pyth{parameters} is a dictonary where additional information can
be passed to the C++ code, e.g., the tolerance for the troubled cell
indicator.
Static methods on \pyth{Model} class passed to \pyth{femdgOperator}:
\enlargethispage{1cm}
\begin{python}
class Model:
    dimRange = r
    # source term (treated explicitly)
    def S_e(t,x,U,DU)      # return $R^r$
    # source term (treated implicitly)
    def S_i(t,x,U,DU)      # return $R^r$
    # diffusion term
    def F_v(t,x,U,DU)      # return $R^{r,d}$
    # advection term
    def F_c(t,x,U):        # return $R^{r,d}$
    # max advection speed used in LLF flux
    def maxLambda(t,x,U,n) # return $>=0$
    # boundary conditions
    boundary = dict(...)
    # for physicality check in troubled cell indicator
    def physical(t,x,U)    # return bool
    # for jump smoothness indicator
    def velocity(t,x,U)    # return $R^d$
    def jump(t,x,U,V)      # return $R^r$
    # for scaling limiter
    Model.lowerBound = # $R^r$
    Model.upperBound = # $R^r$
\end{python}

The final ingredient is the time stepper:
\begin{pythonline}
def femdgStepper(*,order,operator,rkType=None,cfl=0.45,parameters=None)
\end{pythonline}
If \pyth{rkType} is \pyth{None} or \pyth{"default"} then the type of
Runge-Kutta method used will depend on the methods defined on the
\pyth{Model} class used to construct the operator. If no diffusive flux
\pyth{F_v} was defined an explicit RK method is used, if a diffusive
flux but no advective flux \pyth{F_c} is defined an implicit method is
used, while if both fluxes are present an IMEX scheme is employed.
Again the \pyth{parameter} dictonary can be used to set further parameters
read by the underlying C++ code.


\section{C++ code for modal troubled cell indicator}
\label{app:modalind}
The full class for the modal indicator. The class needs to derive from a
templated pure virtual base class
\cpp{Dune::Fem::TroubledCellIndicatorBase<DiscreteFunction>} and override a
single method. The main work is done in the private method
\cpp{smoothnessIndicator} which sets up a least square problem to compute a
smoothness indicator following \cite{Kloeckner}. We assume that the
degrees of freedom \cpp{uEn[i]} are modal. The modal decay is computed for
the first component, i.e.,
\cpp{uEn[0],uEn[dimRange],uEn[2*dimRange],...}:
\begin{c++tiny}
template <class DiscreteFunction>
struct ModalIndicator
: public Dune::Fem::TroubledCellIndicatorBase<DiscreteFunction> {
  using LocalFunctionType = DiscreteFunction::LocalFunctionType;
  ModalIndicator () {}
  double operator()( const DiscreteFunction& U,
                     const LocalFunctionType& uEn) const override {
    double modalInd = smoothnessIndicator( uEn );
    return std::abs( modalInd ) > 1e-14 ? 1. / modalInd : 0.0;
  }

private:
  template <class LF>
  static double smoothnessIndicator(const LF& uLocal) {
    using ONB = Dune::Fem::OrthonormalShapeFunctions<LF::dimDomain>;
    const std::size_t R = LF::dimRange;        // will be using first component
    const std::size_t P = uLocal.order();
    const double area   = uLocal.entity().geometry().volume();
    const double factor = 1/std::sqrt( area ); // scaling (ONB over reference element)

    // compute a 1D moments vector by averaging
    double q[P+1], b2[P+1];
    double f = 0;
    std::size_t k = ONB::size(0); // number of moments of zero degree (is 1)
    q[0] = uLocal[0]*uLocal[0];   // constant part not used
    double l2norm2 = 0;
    for (std::size_t i=1;i<=P;++i) {
      q[i]  = 0;
      b2[i] = 0;
      double nofMoments = ONB::size(i)-k; // number of moments of ith degree
      for (;k<ONB::size(i);++k) { // averaging process
        q[i]    += uLocal[k*R]*uLocal[k*R] / nofMoments;
        l2norm2 += uLocal[k*R]*uLocal[k*R] / nofMoments;
        b2[i]   += pow(1/double(i),2*P)    / nofMoments;
        f       += pow(1/double(i),2*P)    / nofMoments;
      }
    }
    for (std::size_t i=1;i<=P;++i)
      q[i] = std::sqrt( q[i] + l2norm2*b2[i]/f ) / factor;
    double maxQ = std::max( q[P], q[P-1] );

    // find first 'significant' mode
    std::size_t significant = 0;
    for (std::size_t i=P;i>=1;--i) {
      maxQ = std::max(maxQ, q[i]);
      if (maxQ>1e-14) {
        significant = i;
        break;
      }
    }
    if (significant==0) return 1000; // constant, i.e., very smooth indeed
    if (significant==1) return 100;  // linear, not enough info to fit

    // least squares fit to obtain 'smoothness' exponent 's'
    Dune::DynamicMatrix<double> matrix(significant,2);
    Dune::DynamicVector<double> rhs(significant);
    for (std::size_t r=significant; r-->0; ) {
      maxQ = std::max(maxQ, q[r+1]);
      rhs[r]       = std::log( maxQ );
      matrix[r][0] = 1;
      matrix[r][1] = -std::log(double(r+1));
    }
    Dune::FieldMatrix<double,2,2> A;  // LS matrix
    Dune::FieldVector<double,2> b;    // rhs
    for (std::size_t r=0;r<2;++r) {
      for (std::size_t c=0;c<2;++c) {
        A[r][c] = 0;
        for (std::size_t k=0;k<significant;++k)
          A[r][c] += matrix[k][r]*matrix[k][c];
      }
      b[r] = 0;
      for (std::size_t k=0;k<significant;++k)
        b[r] += matrix[k][r]*rhs[k];
    }
    Dune::FieldVector<double,2> x;
    A.solve(x,b);
    return x[1];
  }
};
\end{c++tiny}

\end{appendix}

\end{document}